\documentclass[journal]{IEEEtran}

\usepackage{amsmath,amsfonts,amssymb,mathtools}
\usepackage{multirow}
\usepackage[labelformat=simple]{subcaption}
\usepackage[dvipsnames]{xcolor}

\usepackage[flushleft]{threeparttable}

\usepackage{cite}

\ifCLASSINFOpdf
   \usepackage[pdftex]{graphicx}

\else

   \usepackage[dvips]{graphicx}

\fi

\begin{document}
\title{HDR or SDR? A Subjective and Objective Study of Scaled and Compressed Videos}

\author{Joshua P. Ebenezer,~\IEEEmembership{Student Member,~IEEE,}
        Zaixi Shang,~\IEEEmembership{Student Member,~IEEE,} Yixu Chen,
        Yongjun Wu, Hai Wei, Sriram Sethuraman, Alan C. Bovik,~\IEEEmembership{Fellow,~IEEE} % <-this % stops a space
\thanks{J.P. Ebenezer, Z. Shang, and A.C. Bovik are with the Laboratory for Image and Video Engineering, The University of Texas at Austin, Austin,
TX, 78712, USA e-mail: joshuaebenezer@utexas.edu.}% <-this % stops a space
\thanks{Y. Chen, Y. Wu, H. Wei, and  S. Sethuraman are with Amazon Prime Video.}% <-this % stops a space
\thanks{J.P. Ebenezer and Z. Shang contributed equally to this work}}


\markboth{IEEE Transactions on Image Processing, 2023}%
{Shell \MakeLowercase{\textit{et al.}}: Bare Demo of IEEEtran.cls for IEEE Journals}



\maketitle

\begin{abstract}
 We conducted a large-scale study of human perceptual quality judgments of High Dynamic Range (HDR) and Standard Dynamic Range (SDR) videos subjected to scaling and compression levels and viewed on three different display devices. HDR videos are able to present wider color gamuts, better contrasts, and brighter whites and darker blacks than SDR videos. While conventional expectations are that HDR quality is better than SDR quality, we have found subject preference of HDR versus SDR depends heavily on the display device, as well as on resolution scaling and bitrate. To study this question, we collected more than 23,000 quality ratings from 67 volunteers who watched 356 videos on OLED, QLED, and LCD televisions. Since it is of interest to be able to measure the quality of videos under these scenarios, e.g. to inform decisions regarding scaling, compression, and SDR vs HDR, we tested several well-known full-reference and no-reference video quality models on the new database. Towards advancing progress on this problem, we also developed a novel no-reference model called HDRPatchMAX, that uses both classical and bit-depth sensitive distortion statistics more accurately than existing metrics.  
\end{abstract}

\begin{IEEEkeywords}
High dynamic range, video quality assessment, video compression
\end{IEEEkeywords}

% \fi
\IEEEpeerreviewmaketitle

\section{Introduction}
\IEEEPARstart{H}igh Dynamic Range (HDR) videos have utilize deeper bit-depths to represent brighter and darker luminance with wider color gamuts than Standard Dynamic Range (SDR) videos. To obtain the full benefits of HDR, however, a display must have the technology to accurately represent high contrasts and the extremes of the dynamic range. 
SDR videos are gamma-encoded using the power law described in BT 709~\cite{bt709}. When shown on a TV, these values are decoded and adjusted to the TV's display capabilities using a look-up table. Although SDR standards were originally created for cathode ray televisions having a maximum display brightness of 100 nits, modern display devices can use their entire brightness range (often much greater than 100 nits) to display SDR content, since the digital values of SDR content are relative and not scene-referred. The exact mapping between the SDR digital values and what the television displays are able to display differs across products and is ordinarily proprietary.

Videos following the HDR10 standard, on the other hand, are scene-referred and absolute. The PQ EOTF (used in the HDR10 standard) specifies the absolute luminance that the display must show for a particular digital value. If the absolute luminance value that is required to be shown is greater than the highest luminance value that the TV can display, a tonemapping function called the Electrical-Electrical Transfer Function (EETF) is applied on the HDR content so that clipping does not occur at highlights, while ensuring a smooth roll-off of brightness values at the peak. EETFs differ among televisions and are also usually proprietary.

Due to differences in how SDR and HDR signals are displayed on HDR-capable displays, an SDR version of a content may have a higher average brightness than the HDR version, depending on how each is graded and displayed. For example, a content having a maximum brightness of 200 nits in HDR may be graded in SDR such that the digital value of the maximum brightness is 255. When the HDR and SDR versions are displayed on an HDR-capable device having a peak brightness of 1000 nits, the SDR version may present a peak brightness much larger than 200 nits while the HDR version will be displayed with a peak brightness of 200 nits. The SDR version may therefore appear brighter, but it may also be washed out or oversaturated. Higher peak brightness or higher average brightness are not the only reasons why HDR content can be more appealing than SDR. Indeed, SDR videos may suffer from defects such as saturation, banding, low-contrast, etc., which are less likely to occur in HDR videos. 

In addition to these differences, HDR videos use 10 bit representations (stored in 16 bits), while SDR videos have 8 bit representations. HDR videos require twice the number of bytes that SDR videos of the same content have. Because of this, HDR videos may be more susceptible to compression artifacts.

The tradeoffs between compression, contrast-representation, color-representation, and brightness make the perceptual assessment of HDR and SDR video quality content-dependent and display-dependent. Towards better understanding these tradeoffs, we have conducted a detailed subjective and objective assessment of HDR and SDR videos having the same contents. For the subjective study, we recruited 67 participants who viewed and rated the qualities of 356 HDR and SDR videos of 25 source contents, which were processed by various combinations of scaling and compression using the x265 encoder. We also evaluate objective full-reference (FR) and no-reference (NR) video quality assessment (VQA) on the new subjective database. We also present the design of a new NR VQA model for the task of predicting the quality of both HDR and SDR videos.

\section{Related Work}

To the best of our knowledge, there do not exist any studies that compare the subjective qualities of videos as the dynamic range, resolution, and compression levels are all varied. Existing databases such as LIVE Livestream~\cite{livestream}, LIVE ETRI~\cite{etri}, LIVE YTHFR~\cite{ythfr}, AVT UHD~\cite{avt}, and APV LBMFR~\cite{lbmfr} study the subjective quality of professionally-generated SDR videos under conditions of downsampling, compression, and source distortions. Other datasets including Konvid-1k~\cite{konvid}, YouTube UGC~\cite{ytugc}, and LSVQ~\cite{lsvq} study the quality of SDR user-generated content. UGC databases are typically much larger than those that study the quality of professional-grade content because they can be conducted online via crowdsourcing owing to looser requirements on the display devices, resolution, and bitrate. LIVE HDR~\cite{livehdr}, LIVE AQ HDR~\cite{liveaqhdr}, and APV HDR Sports~\cite{apvhdrsports} are recent databases that study the quality of professionally-created HDR videos that have been downsampled and compressed at various resolutions and bitrates. 
\par 
Each of the above-mentioned databases study the quality of either SDR or HDR videos, but not both., that have been subject to distortions. Here we present the first subjective study that compares the quality of HDR and SDR videos of the same content, that have been processed by downscaling and compression. We conducted the study on a variety of  display devices using different technologies and having differing capabilities.
\par 
While subjective human scores from studies like the one we conducted are considered the gold standards of video quality, conducting such studies is expensive is not scalable. However, objective video quality metrics are designed and trained to automatically predict video quality and can be quite economic and scalable. These fall into two categories: Full-Reference (FR) and No-Reference (NR) models. FR VQA models require as take as input both pristine and distorted videos to measure the quality of the distorted videos. NR metrics only have access to distorted videos when predicting quality, hence designing them is a more challenging problem. NR VQA models are relevant for video source inspection as well as when measuring quality with no available source video.
\par    
PSNR measures the peak signal to noise ratio between a reference frame and a distorted version of the same frame. SSIM~\cite{ssim} incorporates luminance, contrast, and structure features to predict the quality of distorted images. VMAF~\cite{vmaf} models the statistics of the wavelet coefficients of video frames, as well as the detail losses from distortions. SpEED~\cite{speed} measures the difference in entropy of bandpass coefficients of reference and distorted videos. STRRED~\cite{strred} models the statistics of space-time video wavelet coefficients altered by distortions. STGREED~\cite{greed} measures differences in temporal and spatial entropy arising from distortions to model the quality of videos having varying frame rates and bitrates.

BRISQUE~\cite{brisque}, VBLIINDS~\cite{vbliinds}, VIDEVAL~\cite{videval}, RAPIQUE~\cite{rapique}, ChipQA~\cite{chipqa}, HDR ChipQA~\cite{hdrchipqa} and NIQE~\cite{niqe} are NR video quality metrics that rely on neurostatistical models of visual perception. Pristine videos are known to follow certain regular statistics when processed using visual neural models. Distortions predictably alter the statistics of perceptually processed videos, allowing for the design of accurate VQA models. RAPIQUE combines features developed under these models with (semantic) video features provided by a pre-trained deep network. TLVQM~\cite{tlvqm} explicitly models common distortions such as compression, blur, and flicker, using a variety of spatial and temporal filters and heuristics. 
\section{Details of Subjective Study}
The study was conducted on 356 videos shown to 67 subjects. The videos were generated from a set of 25 unique and pristine contents also converted to SDR, all subjected to combinations of downscaling and HEVC compression. Three different television technologies were used to conduct the study.

\subsection{Source Sequences}

The 25 source sequences can be divided into 4 groups: Video on Demand (VoD), Live Sports, HDR Demo Videos, and SJTU videos. Three of the source sequences were ``anchor" sequences taken from prior HDR VQA databases (LIVE HDR, LIVE AQ HDR, and APV HDR Sports) to calibrate and combine their data with the present database. All of the videos are represented in the BT2020~\cite{bt2020} color gamut and were quantized using the SMPTE ST2084~\cite{pq} Opto-Electronic Transfer Function (OETF), also known as the Perceptual Quantizer (PQ). All of the source sequences have durations in the range 7s and 10s, with static metadata conforming to the HDR10 standard. The video categories are described in detail as follows:
\subsubsection{VoD}
These 7 videos were professionally captured and graded for VoD streaming services. The HDR and SDR versions of these videos were prepared by Amazon Studios. One of the contents used in this category is an ``anchor" video from the LIVE AQ HDR database. The SDR versions were created and manually graded with creative intent by professional graders.

\subsubsection{Live Sports}
The 4 Live Sports videos were captured professionally by broadcasters at stadiums hosting live matches of Soccer and Tennis. One content from this category is an ``anchor" video from the APV HDR Sports database. Due to the low-latency requirements of live broadcasts, these videos were graded using preset Lookup Tables (LUTs) for both HDR and SDR formats. The LUT used for the HDR grading is a proprietary Amazon LUT, while the LUT used for the HDR to SDR conversion is the open-source NBC LUT. All the videos were originally in YUV422 10-bit format and were converted to the limited YUV420 10-bit format. 

\subsubsection{HDR Demo Videos}
These are a set of 8 open-source videos collected from 4kmedia.org. The videos were created and graded by television manufacturers to showcase the capabilities of HDR over SDR, and hence these contents have a high degree of contrast and colorfulness in order to be eye-catching. The source videos are in the limited YUV420 10-bit format. They were converted to their SDR versions using the NBC LUT.

\subsubsection{SJTU videos}
Six source contents belong to this category. These contents were taken from the open-source SJTU HDR Video Sequence Dataset. They were recorded using a Sony F65 camera and graded using the S-Gamut LUT. The videos are in the limited YUV420 10-bit format and were also converted to their SDR versions using the NBC LUT. All of the source contents in this category are also present in the LIVE HDR database, although the distorted versions differ.

\subsection{Content Descriptors}

\begin{figure}
     \centering
     \begin{subfigure}[b]{0.22\textwidth}
         \centering
         \includegraphics[width=\textwidth]{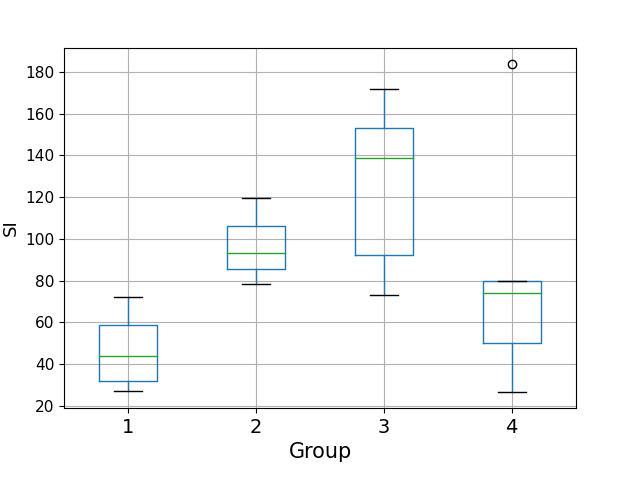}
         \caption{SI}
         \label{fig:si}
     \end{subfigure}
     \hfill
     \begin{subfigure}[b]{0.22\textwidth}
         \centering
         \includegraphics[width=\textwidth]{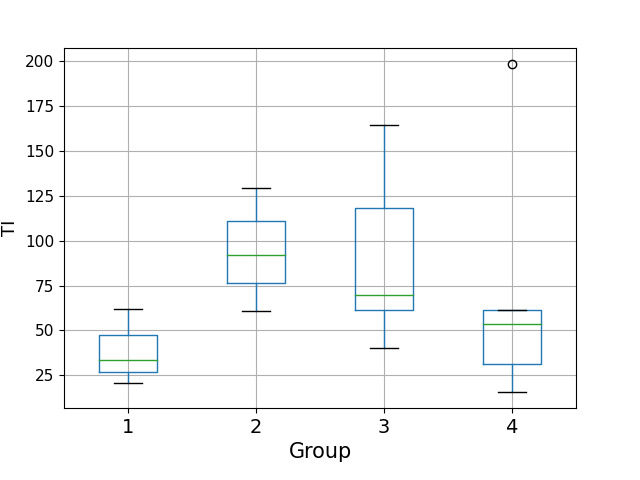}
         \caption{TI}
         \label{fig:ti}
     \end{subfigure} \\
     \begin{subfigure}[b]{0.22\textwidth}
         \centering
         \includegraphics[width=\textwidth]{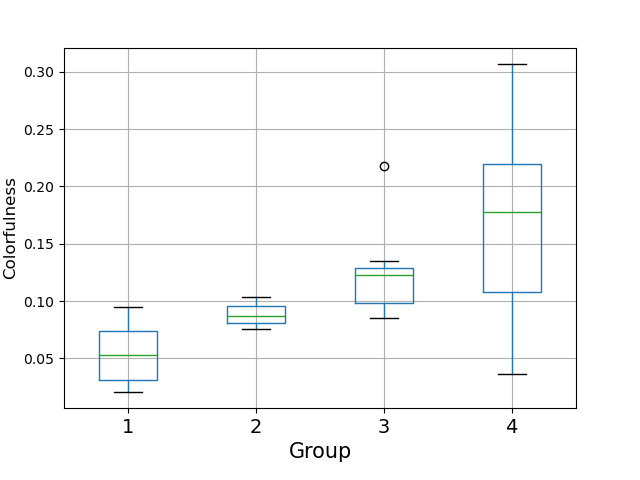}
         \caption{Colorfulness}
         \label{fig:cf}
     \end{subfigure}
     \begin{subfigure}[b]{0.22\textwidth}
         \centering
         \includegraphics[width=\textwidth]{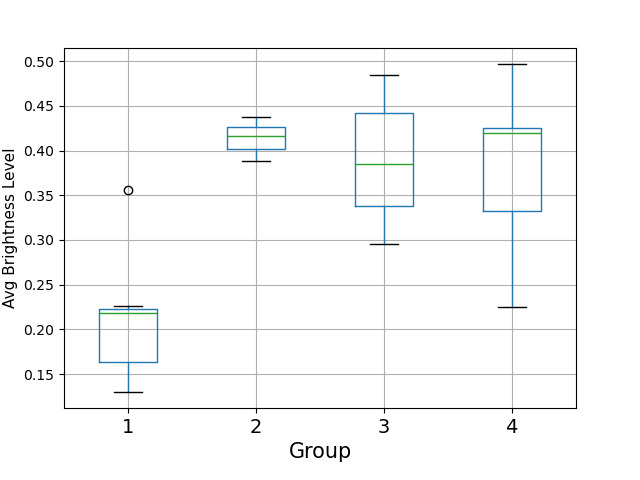}
         \caption{Avg. Luminance}
         \label{fig:abl}
     \end{subfigure}     
        \caption{Content Descriptors, where 1=VoD videos, 2=Live Sports videos, 3=HDR demo videos, and 4=SJTU videos. See text for details.}
        \label{fig:group1_contentdescriptors}
\end{figure}

The Spatial Information (SI), Temporal Information (TI), Colorfulness, and Average Luminance Level were computed for each video sequence and plotted by their groups in Fig.~\ref{fig:group1_contentdescriptors}. 
The VoD contents can be characterized as having lower average brightness levels, lower SI, lower TI, and a lower colorfulness index than the other categories. 
The Live content has high TI.
The Demo Videos have a high level of contrast, colorfulness, and brightness since they were designed to showcase HDR's capabilities.
The SJTU videos include many different scenes and hence have a high degree of variability. Their SI and TI are lower than that of Groups 2 and 3, although one video of fireworks was an outlier in this group for all four content descriptors. 

\subsection{Processing of Source Sequences}

The VoD contents include studio-graded HDR and SDR source versions, while the other content categories are pristine HDR versions which were also converted to SDR versions using the NBC LUT. The HDR and SDR versions were encoded using the bitrate-capped Constant Rate Factor (CRF) method of the x265 encoder. The maximum bitrates (maxrate), CRFs, and buffer sizes for each resolution are listed in Table~\ref{tab:maxrate_res}.

\begin{table*}
\caption{}
\begin{center} 
  \begin{tabular}{|c|c|c|c|c|}
  \hline
   Resolution & CRF for HDR & CRF for SDR & Maxrate (kbps) & Buf. Size (kb)  \\
   \hline
   540p & 25 & 30 & 800 & 2000  \\
   \hline
720p & 17 & 24.5 & 3000 & 6000  \\
\hline
1080p & 20 & 27 & 3000 & 8000 \\
\hline
2160p & 25 & 28 & 6000 & 15000 \\
\hline
1440p & 15 & 19 & 10000 & 20000 \\
\hline
2160p & 17.5 & 22 & 15000 & 22500 \\
\hline
\end{tabular}
\label{tab:maxrate_res}
\end{center}
\end{table*}

Nine HDR videos of the EPL6 content (from the Live Sports category) were taken from the APV HDR Sports database, nine HDR videos of the NightTraffic content (from the SJTU category were taken from the LIVE HDR database, and nine HDR videos of the BTFB-01h05m40s content were taken from the LIVE AQ HDR database. Each group of nine videos included the pristine version as well as eight compressed versions. These 27 ``anchor" videos were used to estimate a mapping between the scores from prior HDR databases to the scores in the new HDR vs SDR database. Each of the anchor contents also had 6 compressed SDR versions at the maxrates and resolutions shown in Table~\ref{tab:maxrate_res}, as well as a pristine SDR version. Hence each anchor content was associated with 16 video sequences each (9 HDR versions and 7 SDR versions).

The remaining 22 source contents each were associated a pristine HDR version, a pristine SDR version, six compressed HDR versions and six compressed SDR versions, for a total of 308 videos. These videos, combined with those from the anchor content, yield a total of 356 videos in the new database. {There are thus a total of 181 HDR videos and 175 SDR videos in the database.}

\subsection{Display Devices}

Three televisions were selected for the study: the 65" Samsung S95T (TV1), 65" Samsung Q90 (TV2), and the 55" Amazon Fire TV (TV3). All of the televisions are capable of receiving and displaying HDR10 content. Peak luminance refers to the maximum brightness a television can get when displaying HDR. Different televisions advertise different ``peak luminance" capabilities, but they may only be able to generate the advertised value in a small portion of the screen for a short duration. This is done to prevent damage to the display and to reduce power consumption, and is referred to as Auto Brightness Limiting. We refer to the the peak luminance as the instantaneous brightness of a white rectangle displayed on an area covering 2\% of the screen, {as reported by \cite{rtings}.}
\par The Samsung S95 has a quantum dot organic light-emitting diode (QD-OLED) display. Each pixel emits its own light, and hence the device can show very high contrasts. It has 86.93\% coverage of the BT 2020 color space and a peak luminance of 1028 cd/m\textsuperscript{2}. The Samsung Q90T is a Quantum Dot display with a Vertical Alignment (VA) LED backlight. Quantum Dots emit red and green colors with high accuracy, and can hence produce more vivid colors than standard LED TVs. The Samsung Q90 has 67.24\% coverage of the BT 2020 color space and a peak luminance of 1170 cd/m\textsuperscript{2}. The Amazon Fire TV is an entry-level VA LED TV with a 54.25\% coverage of the Rec 2020 color gamut and a peak luminance of 230 cd/m\textsuperscript{2}. The Samsung Q90T and the Amazon Fire TV have the full array local dimming (FALD) feature that dims the backlight brightness in areas of the screen that are meant to be darker in order to increase the contrast. However, due to the backlight, brightness can still ``bleed" from brighter areas of the screen to darker areas, which can affect contrast. The Samsung S95, on the other hand, is an OLED display, hence each pixel can be controlled individually for better contrasts than FALD can allow. However, the Samsung Q90T can achieve higher peak brightness values than the S95 because of the presence of the LED backlight as well as the Quantum Dots, which amplify light. 

A Windows PC with an NVIDIA 3090 GPU running the Windows 10 Operating System was used to drive the televisions via a HDMI 2.1 cable. The PC and the televisions had HDR enabled. The screen resolutions were set at 3840x2160 and the refresh rate was set at 30 Hz. The VLC media player was used for video playback.

\subsection{Subjects}

A total of 67 students at the University of Texas at Austin volunteered to participate in the human study. All the subjects were between the ages of 20 and 28. Approximately two-thirds of the subjects were male, and the remaining third identified as  female. A demographic survey revealed that 73\% of the subjects identified as Asian, 20\% identified as White, and 7\% identified as Black. 

Among these, 22 subjects were assigned to watch TV1, 21 were assigned to watch TV2, and 24 were assigned to watch TV3. None of the subjects were told about the other TVs or the nature of the study in order to eliminate biases. All the subjects passed the Ishihara test for color-blindness and the Snellen test for visual acuity when wearing their corrective lenses (if needed).

\subsection{Subjective Testing Design}

We employed a Single Stimulus method for the study, as described in ITU-R BT 500.13~\cite{bt500}. Each video was shown once to each subjects and a quality score was collected from the subject immediately after the video was shown. The videos were displayed in random order and the reference and distorted videos were not identified or given different treatment. Videos of the same content were not allowed to be adjacent in the viewing order, in order to reduce memory biases. Each quality score was collected using an invisible integer scale from 1-100 with a continuous slider that had 5 verbal markers: ``Poor," ``Bad," ``Fair," ``Good," and ``Excellent." The slider was operated by a mouse. The study was divided into two sessions of approximately 40 minutes each, and the two sessions were separated by at least 24 hours to reduce viewer fatigue.

Before each test session began, a training session was conducted whereby the subject was familiarized with the setup using videos that are not a part of the database. A set of six videos of the same content at varying levels of compression were shown to each subject, three being in HDR and three being in SDR, such that the quality range present in the database was fairly represented. Subjects were shown how to use the scoring mechanism. During the training session, instructions were given on how to rate the videos based on their subjective assessment of the quality, while avoiding judgments on the aesthetic content. No other instructions or details about the study were given to avoid biasing the participants.

\section{Subjective Analysis}

\subsection{Internal Correlation}

The internal correlations for the three groups were calculated as follows. Let the score given by subject $i$ for video $j$ on TV $t$ be given by $u_{ijt}$. The $Z$ score is computed as 
\begin{equation}
    Z_{ijt} = \frac{u_{ijt}-\frac{\Sigma_j u_{ijt}}{N}}{\sqrt{\frac{\Sigma_j (u_{ijt}-\Sigma_j u_{ijt})^2}{N}}}.
\end{equation}
The subjects were randomly divided into two equal groups and the average $Z$ score computed for each video across all subjects in that group. The correlation between the scores provided by these groups for all the videos was computed over 100 trails with different random groupings. The median inter-subject correlation for viewers of TV1 was found to be 0.95, for TV2 was 0.94, and for TV3 was 0.93. These data indicate a high degree of internal consistency and reliability.   

\subsection{Calculation of MOS}

The Mean Opinion Scores (MOS) were obtained using the Maximum Likelihood Estimation method proposed in ITU 910~\cite{ITU910}. The MOS is modelled as a random variable

\begin{equation}
    U_{ijn} = \Psi_{jn} + \Delta_i + \nu_i X
    \label{eq:sureal}
\end{equation}

where $\Psi_{jn}$ is the true quality of video $j$ viewed on TV$n$, $\Delta_i$ is the bias of subject $i$, $\nu_i$ represents the inconsistency of subject $i$, and $X\sim N(0,1)$ are i.i.d. Gaussian random variables. Given the scores $u_{ijn}$, the true score for each video on each television is estimated by treating $\Psi_{jn},\Delta_i, \nu_i$ as free parameters that are solved so that the model in (\ref{eq:sureal}) is the best fit to the observed MOS. Specifically, $\Psi_{jt}$ is by maximizing the log-likelihood of the observations using the Newton-Raphson solver.

The Differential Mean Opinion Scores (DMOS) were calculated between the distorted and reference videos by finding the differences in MOS as follows:
\begin{equation}
    DMOS_{jt} = \Psi_{jt}-\Psi_{j_0 t},
\end{equation}
where $j_0$ is the index of the reference video corresponding to video $j$ viewed on television TV$n$. The MOS and DMOS are thus computed for each video and separately for each television.

\subsection{Analysis of Scores}

\begin{figure*}
     \centering
          \begin{subfigure}[b]{0.24\textwidth}
         \centering
         \includegraphics[width=\textwidth]{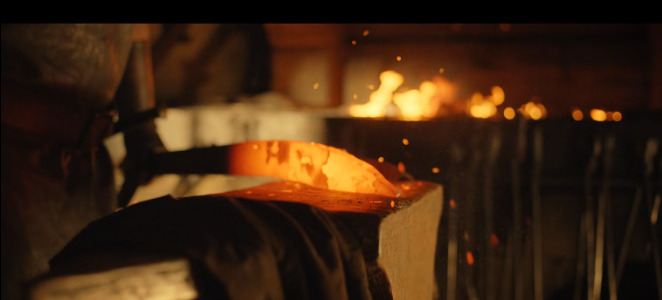}
         \caption{Screenshot from video ``Forge".}
         \label{fig:ltum}
     \end{subfigure}
     \begin{subfigure}[b]{0.24\textwidth}
         \centering
         \includegraphics[width=\textwidth]{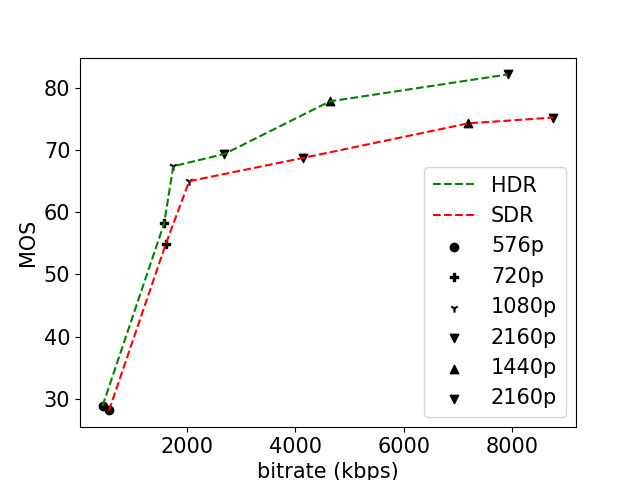}
         \caption{MOS vs bitrate for TV1}
         \label{fig:tv1_ltum}
     \end{subfigure}
     \hfill
     \begin{subfigure}[b]{0.24\textwidth}
         \centering
         \includegraphics[width=\textwidth]{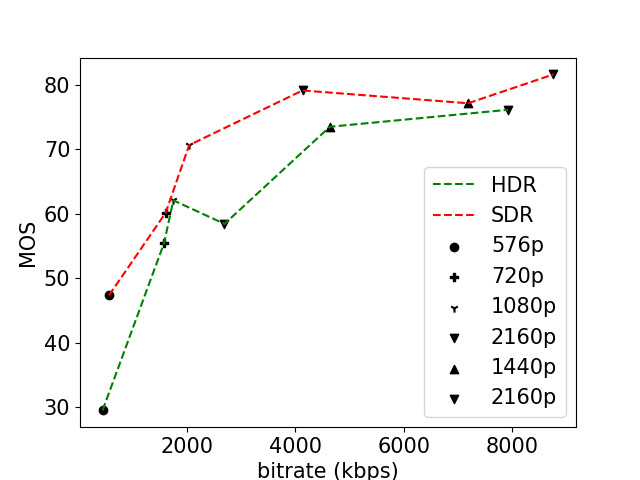}
         \caption{MOS vs bitrate for TV2}
         \label{fig:tv2_ltum}
     \end{subfigure}
     \hfill
     \begin{subfigure}[b]{0.24\textwidth}
         \centering
         \includegraphics[width=\textwidth]{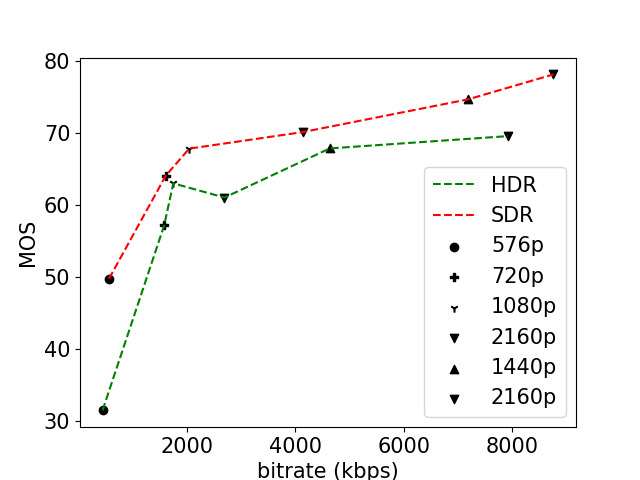}
         \caption{MOS vs bitrate for TV3}
         \label{fig:tv3_ltum}
     \end{subfigure}
        \caption{MOS vs bitrate plots for the three tested televisions on the ``Forge" video.}
        \label{fig:group1_mos}
\end{figure*}

\begin{figure*}
     \centering
          \begin{subfigure}[b]{0.24\textwidth}
         \centering
         \includegraphics[width=\textwidth]{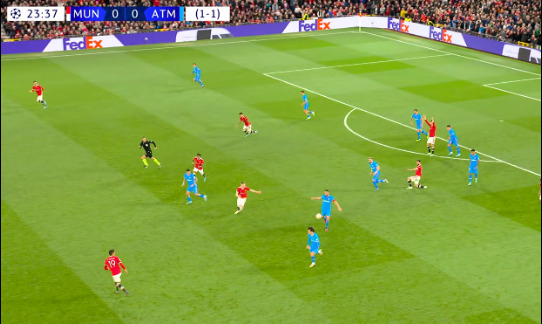}
         \caption{Screenshot from video ``EPL".}
         \label{fig:tv1_epl_Screenshot}
     \end{subfigure}
     \hfill
     \begin{subfigure}[b]{0.24\textwidth}
         \centering
         \includegraphics[width=\textwidth]{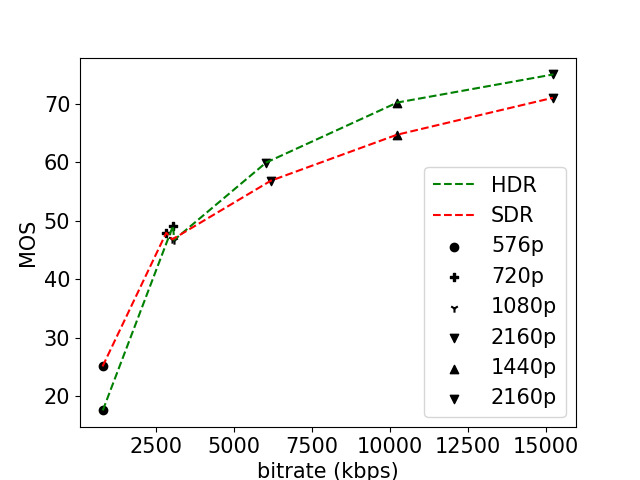}
         \caption{MOS vs bitrate for TV1}
         \label{fig:tv1_epl}
     \end{subfigure}
     \hfill
     \begin{subfigure}[b]{0.24\textwidth}
         \centering
         \includegraphics[width=\textwidth]{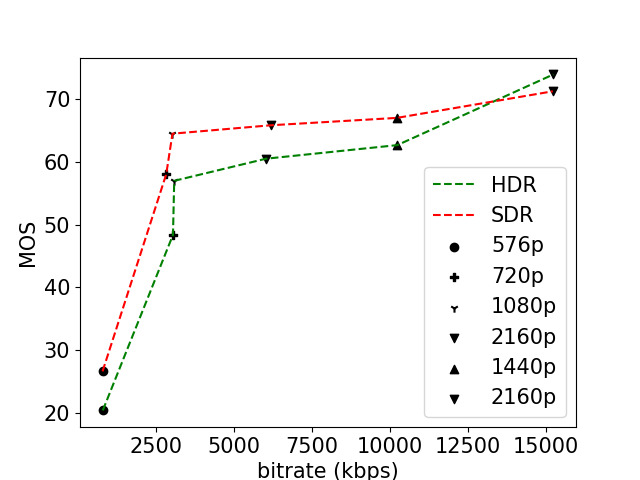}
         \caption{MOS vs bitrate for TV2}
         \label{fig:tv2_epl}
     \end{subfigure}
     \hfill
     \begin{subfigure}[b]{0.24\textwidth}
         \centering
         \includegraphics[width=\textwidth]{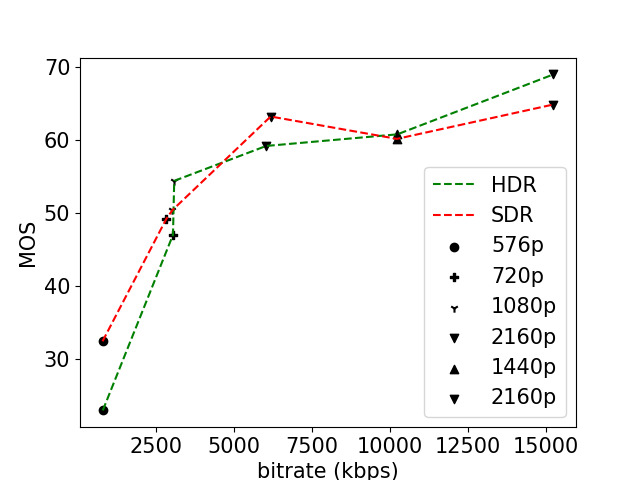}
         \caption{MOS vs bitrate for TV3}
         \label{fig:tv3_epl}
     \end{subfigure}
        \caption{MOS vs bitrate plots for the three tested televisions on the ``EPL" video.}
        \label{fig:group2_mos}
\end{figure*}

\begin{figure*}
     \centering
          \begin{subfigure}[b]{0.24\textwidth}
         \centering
         \includegraphics[width=\textwidth]{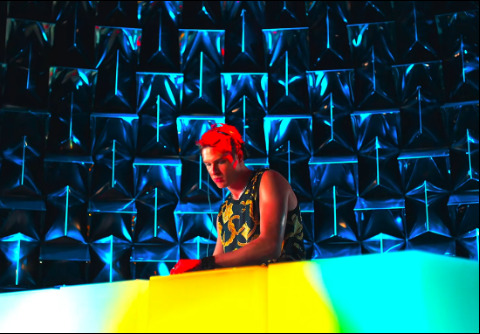}
         \caption{Screenshot from video ``ColorDJ".}
         \label{fig:tv1_Screenshot}
     \end{subfigure}
     \hfill
     \begin{subfigure}[b]{0.24\textwidth}
         \centering
         \includegraphics[width=\textwidth]{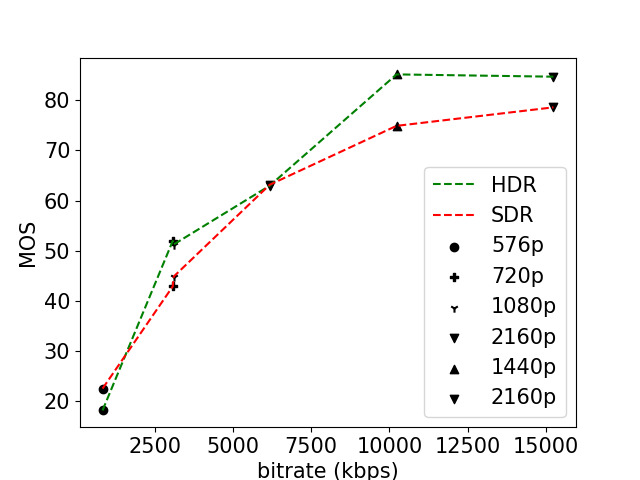}
         \caption{MOS vs bitrate for TV1}
         \label{fig:tv1_group3}
     \end{subfigure}
     \hfill
     \begin{subfigure}[b]{0.24\textwidth}
         \centering
         \includegraphics[width=\textwidth]{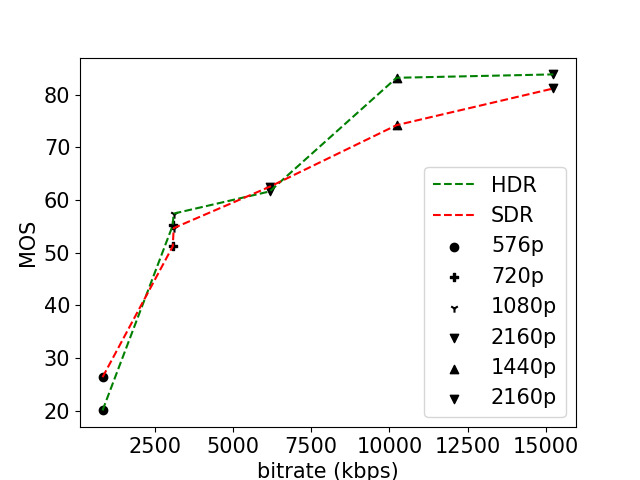}
         \caption{MOS vs bitrate for TV2}
         \label{fig:tv2_group3}
     \end{subfigure}
     \hfill
     \begin{subfigure}[b]{0.24\textwidth}
         \centering
         \includegraphics[width=\textwidth]{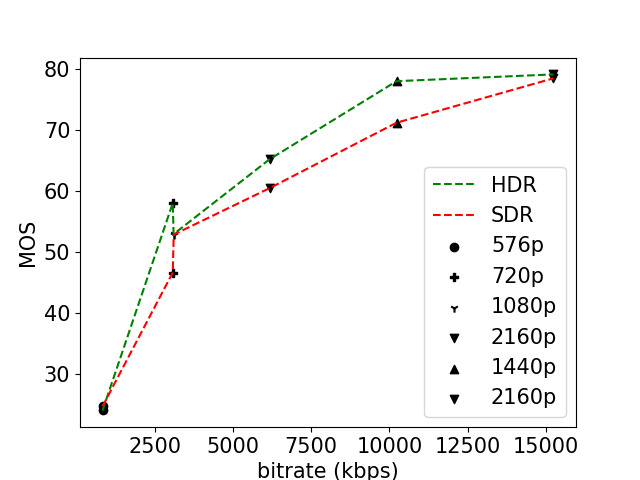}
         \caption{MOS vs bitrate for TV3}
         \label{fig:tv3_group3}
     \end{subfigure}
        \caption{MOS vs bitrate plots for the three tested televisions on the ``ColorDJ" video.}
        \label{fig:group3_mos}
\end{figure*}

\begin{figure*}
     \centering
          \begin{subfigure}[b]{0.24\textwidth}
         \centering
         \includegraphics[width=\textwidth]{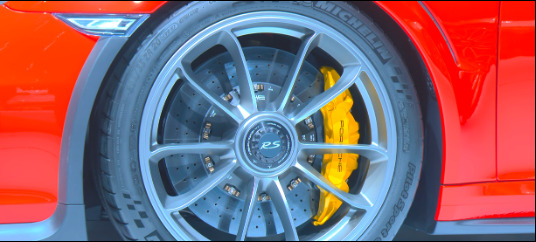}
         \caption{Screenshot from video ``Porsche".}
         \label{fig:porsche}
     \end{subfigure}
     \hfill
     \begin{subfigure}[b]{0.24\textwidth}
         \centering
         \includegraphics[width=\textwidth]{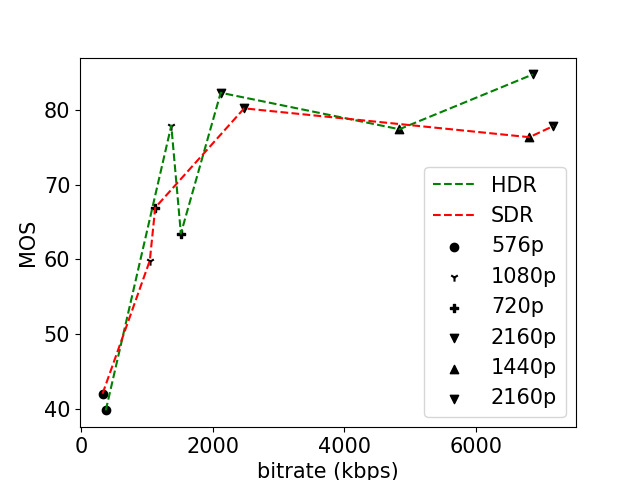}
         \caption{TV1}
         \label{fig:tv1_group4}
     \end{subfigure}
     \hfill
     \begin{subfigure}[b]{0.24\textwidth}
         \centering
         \includegraphics[width=\textwidth]{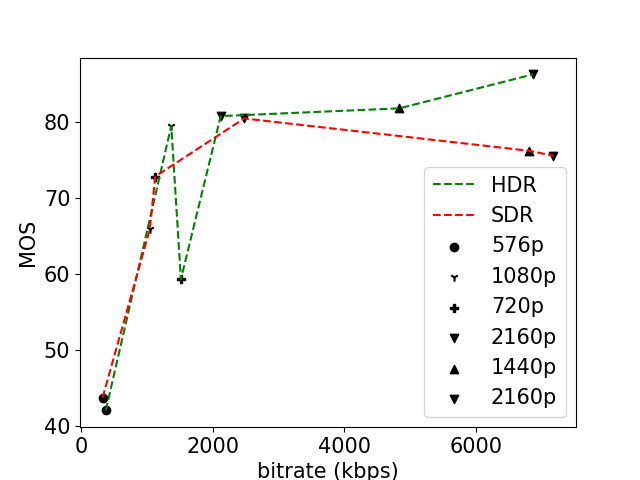}
         \caption{TV2}
         \label{fig:tv2_group4}
     \end{subfigure}
     \hfill
     \begin{subfigure}[b]{0.24\textwidth}
         \centering
         \includegraphics[width=\textwidth]{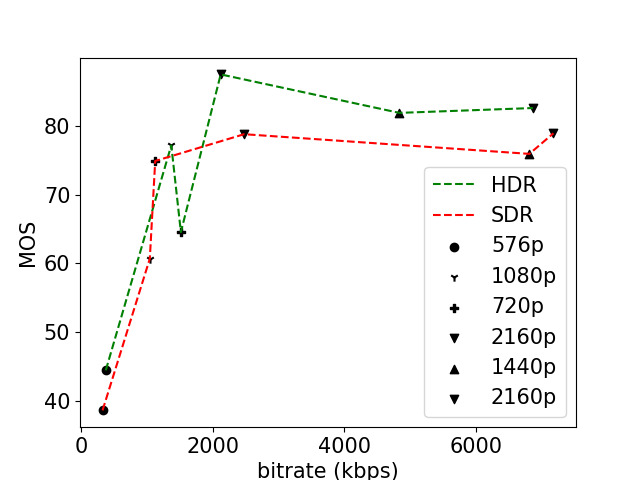}
         \caption{TV3}
         \label{fig:tv3_group4}
     \end{subfigure}
        \caption{MOS vs bitrate plots for the three tested televisions on the ``Porsche" video.}
        \label{fig:group4_mos}
\end{figure*}

The MOS of 20 videos of 4 contents (1 content from each source group and 5 videos per content) are plotted against bitrate in Figs. \ref{fig:group1_mos}, \ref{fig:group2_mos}, \ref{fig:group3_mos}, and \ref{fig:group4_mos}. The Forge video is from group 1, EPL is from group 2, ColorDJ is from group 3, and Porsche is from group 4. Screenshots from the SDR versions of the source contents are shown in the first column. The points on each line correspond to 540p, 720p, 1080p, 1440p, 2160p, and 2160p in increasing order, following Table \ref{tab:maxrate_res}. 

The Forge video shows a sword being forged in a smithy. The light from the hot iron contrasts strongly with the darkness and shadows around it and in the background. The HDR version of the Forge video is rated higher than the SDR version on TV1, since details are clearer in the HDR version. However, on TV2 and TV3, due to their reduced ability to display contrasts, higher compression of the HDR video offsets the relative quality gain from the increased contrast, and the SDR version was thus rated as better. The HDR version appears darker and dimmer than the SDR version, due to the way in which HDR is displayed differently from SDR (as discussed in the Introduction) as well as the way in which the video was graded. The lower average brightness may also contribute to the lower perceived quality of the HDR version of the video content. The lower average brightness of the HDR version, the reduced contrast of TV2 and TV3, and the high spatial complexity of the scene, may also explain why, at higher bitrates, the SDR version was still rated higher. Similar observations can be made about the EPL video taken from group 2, where the HDR version was rated better on TV1, while the SDR version was rated better at lower bitrates on TV2 and TV3. At higher bitrates, the HDR version was rated higher, as compressive artifacts had less visual impact, despite the high temporal complexity of the soccer scene.

The SDR version of the ColorDJ video suffers from over-saturation and overexposure due to the wide range of colors and brightnesses present. The HDR version, on the other hand, does not exhibit over-saturation or overexposure because of its greater bit-depth and wider color gamut. This may explain why the HDR version was generally rated better on all the televisions.

The Porsche video has a bright red paint on the car which is accurately represented in HDR, but looks saturated in SDR. The video is also not spatially or temporally complex, which may be because of the superior ability of HDR to represent contrasts and bright colors, outweighing the effects of compression. However, there is a sharp drop in the MOS vs bitrate curve for the 720p HDR version of the video, encoded at 1515 kbps. The encoder made the decision to encode the 1080p version at 1300 kbps, yet the 1080p version is still rated higher than the 720p version. This is likely because rescaling artifacts are more prominent than the compressive artifacts on this content due to its low spatial complexity.

\begin{figure*}
    \centering
     \begin{subfigure}[b]{0.3\textwidth}
         \centering
         \includegraphics[width=\textwidth]{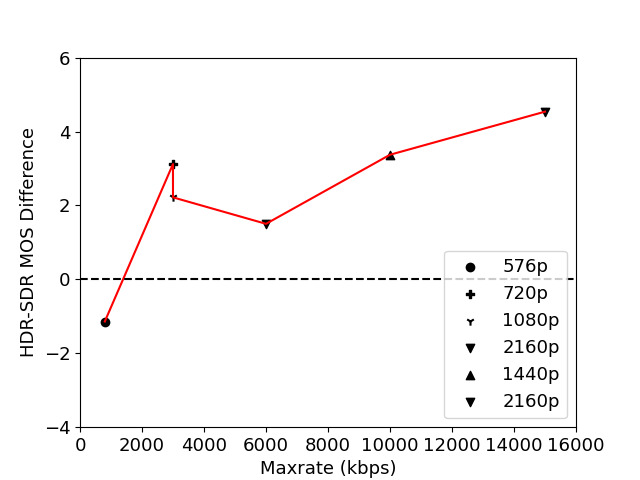}
         \caption{MOS vs bitrate for TV1}
         \label{fig:tv1_diff}
     \end{subfigure}
          \hfill
     \begin{subfigure}[b]{0.3\textwidth}
         \centering
         \includegraphics[width=\textwidth]{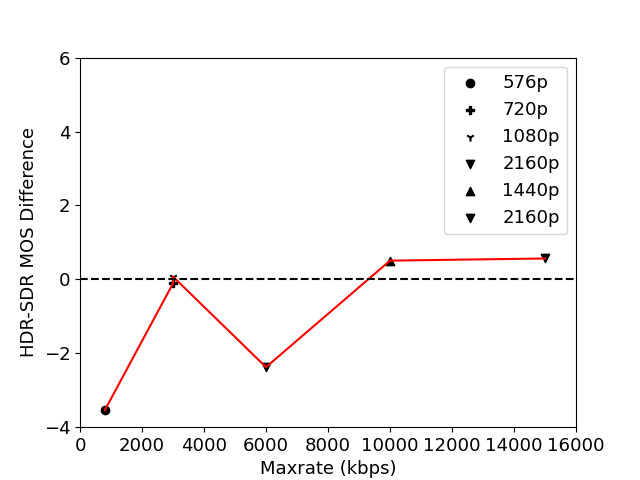}
         \caption{MOS vs bitrate for TV2}
         \label{fig:tv2_diff}
     \end{subfigure}
          \hfill
     \begin{subfigure}[b]{0.3\textwidth}
         \centering
         \includegraphics[width=\textwidth]{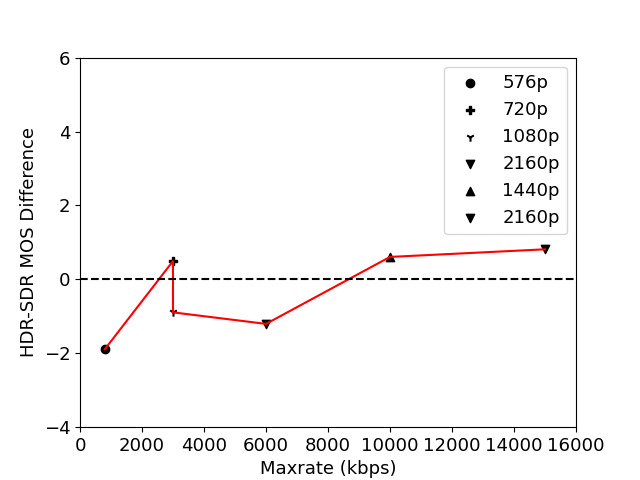}
         \caption{MOS vs bitrate for TV3}
         \label{fig:tv3_diff}
     \end{subfigure}
     \caption{Average difference between MOS of corresponding HDR and SDR videos vs maxrate for different televisions.}
\label{fig:tvdiff}
\end{figure*}

The average difference between the MOS of the HDR and SDR versions of all the videos are plotted against the maxrate of each television in Fig.~\ref{fig:tvdiff}. As may be seen, at lower bitrates SDR was rated higher than HDR, while at higher bitrates the difference becomes positive. The drops in the curve of 1080p content at maxrate 3000 kbps and of 2160p at maxrate 6000 kbps are indicative of how for an optimal bitrate ladder, these resolutions should be encoded at higher bitrates. It is important to note that these compression levels were included in the study to represent a wide range of quality, and not to represent an optimal bitrate ladder. For TV1, HDR was better than SDR at even low bitrates (3000 kbps) and the difference in quality increased to 5 MOS units at the highest bitrate. For TV2 and TV3, SDR quality was better than HDR quality at low bitrates but the differences decreased and became slightly positive at higher bitrates. This also indicates how the capabilities of the display devices strongly influence the quality of HDR relative to that of SDR content. 

\subsection{Combining Databases}
\label{section:combiningdata}

The scores given to the 27 anchor videos from the LIVE HDR, LIVE AQ HDR, and the APV HDR Sports VQA datasets were used to map scores from those databases to the current database. A logistic function was fitted to map the scores assigned to the anchor videos in those databases to the scores of the same videos in the LIVE HDRvsSDR database for each television. The logistic function is

\begin{equation}
    f(x) = \frac{a-b}{(1 + \exp \frac{-(x - c)}{s})} + b,
\end{equation}
 where $x$ are the scores of the videos in prior databases, $f(x)$ is the mapping to scores in the LIVE HDRvsSDR database for a particular television, and $a,b,c$ and $s$ were separately solved for on each database using the anchor videos for that database. Since each television and each (prior) database was used to generate a different fitting function, a total of nine functions were derived from the data. The functions that map scores from the LIVE HDR database and the LIVE AQ HDR database to the scores obtained from the three televisions used in the LIVE HDRvsSDR database are plotted in Figs. 12 and 13 in supplementary material, respectively. The mappings and results for the APV HDR Sports database cannot be shown, for proprietary reasons.

 Deriving these functions from the scores of the anchor videos enables the merging of the three LIVE databases into a single, large-scale database of 1066 videos that were collected in a controlled laboratory environment.                                                     

\section{Objective Assessment}

\subsection{Full Reference Video Quality Assessment}

We tested the Peak Signal to Noise Ratio (PSNR), the Structural Similarity Index Measure (SSIM)~\cite{ssim},  Multi-Scale SSIM (MS-SSIM)~\cite{msssim}, Video Multimethod Assessment Fusion (VMAF)~\cite{vmaf}, Spatio-Temporal Reduced Reference Entropic Differences (STRRED)~\cite{strred},  Spatial Efficient Entropic Differencing (SpEED)~\cite{speed}, and the Space-Time GeneRalized Entropic Difference (STGREED)~\cite{greed} video quality models on the newly created database. 
We evaluated the FR VQA algorithms by computing the Spearman's Rank Ordered Correlation Coefficient (SRCC) between the scores predicted by the algorithms and the ground truth DMOS. We also fit the predicted scores to the DMOS using a logistic function
\begin{equation}
l(s) =  \frac{\beta_1-\beta_2}{ 1 + \exp(-\frac{(x - \beta_3)}{ \beta_4}) + \beta_5},
\label{eq:logistic}
\end{equation}

and then computed Pearson's Linear Correlation Coefficient (LCC) and the Root Mean Square Error (RMSE) between the fitted scores and the DMOS, following standard practice~\cite{logistic}. 
The results are presented in Table~\ref{tab:fr_metrics} .

\begin{figure*}
     \centering
          \begin{subfigure}[b]{0.3\textwidth}
         \centering
         \includegraphics[width=\textwidth]{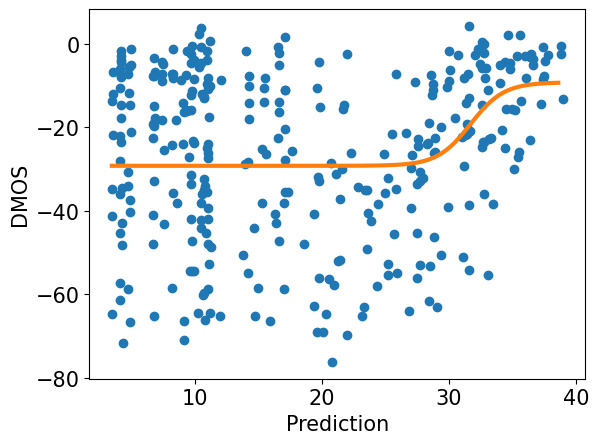}
         \caption{PSNR}
         \label{fig:psnr}
     \end{subfigure}
     \hfill
     \begin{subfigure}[b]{0.3\textwidth}
         \centering
         \includegraphics[width=\textwidth]{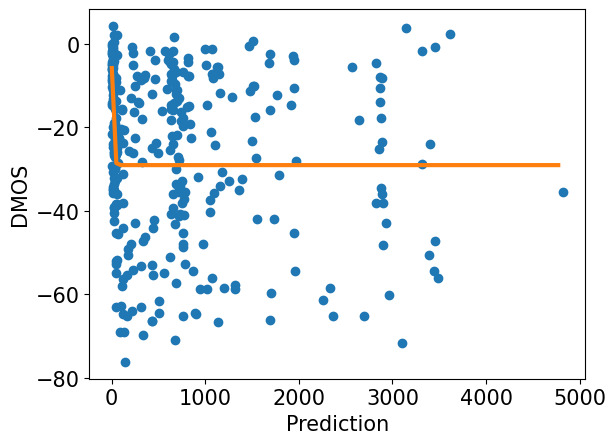}
         \caption{STRRED}
         \label{fig:strred}
     \end{subfigure}
\hfill
     \begin{subfigure}[b]{0.3\textwidth}
         \centering
         \includegraphics[width=\textwidth]{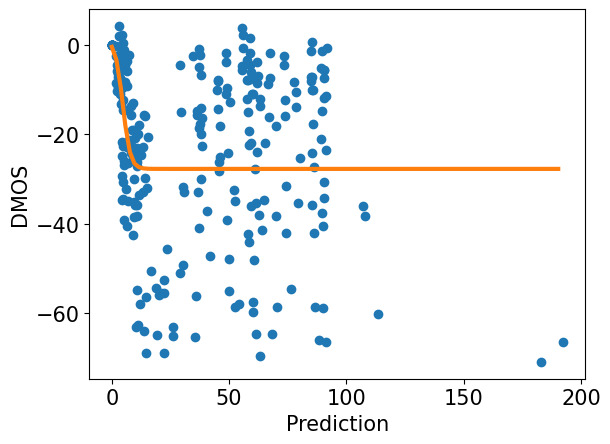}
         \caption{SpEED}
         \label{fig:speed}
     \end{subfigure}
     \\          
     \begin{subfigure}[b]{0.3\textwidth}
         \centering
         \includegraphics[width=\textwidth]{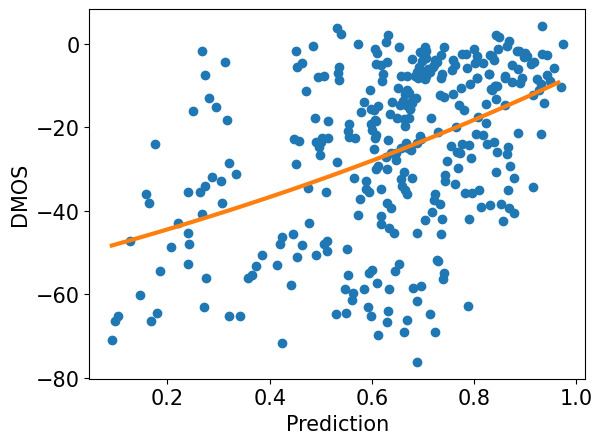}
         \caption{MS-SSIM}
         \label{fig:msssim}
     \end{subfigure}
     \hfill
          \begin{subfigure}[b]{0.3\textwidth}
         \centering
         \includegraphics[width=\textwidth]{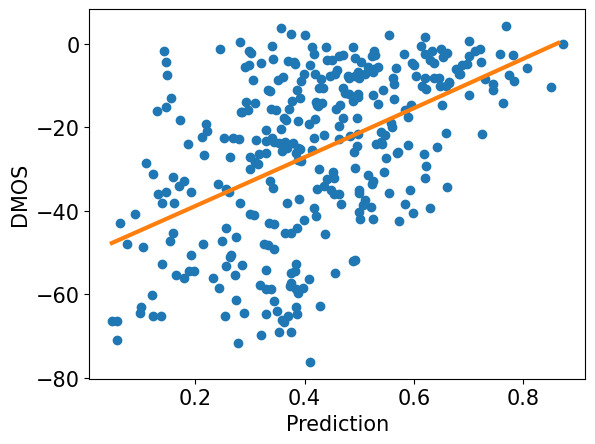}
         \caption{SSIM}
         \label{fig:ssim}
     \end{subfigure}
        \hfill
               \begin{subfigure}[b]{0.3\textwidth}
         \centering
         \includegraphics[width=\textwidth]{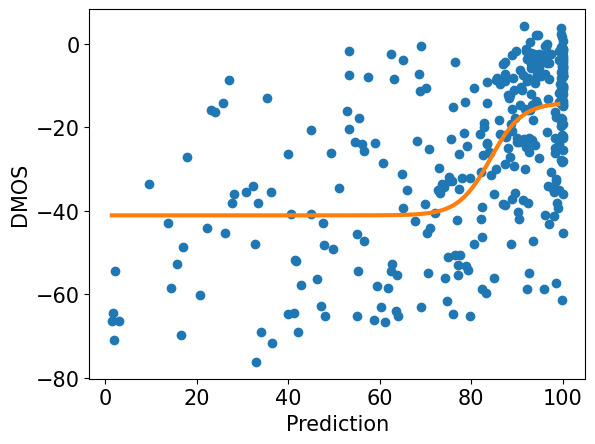}
         \caption{VMAF}
         \label{fig:vmaf}
     \end{subfigure}
\caption{Scatter plots of DMOS for videos shown in TV1 against FR VQA predictions with logistic fit in orange.}
        \label{fig:frpredictions}
\end{figure*}

As can be seen, the FR metrics did not perform well on the task, with VMAF achieving the highest SRCC of 0.56 on TV1, 0.55 on TV2, and 0.59 on TV3. In addition, since none of these models incorporate modelling of the display device, their predictions were the same on all three televisions, except for STGREED which requires an SVR to be trained separately for each TV.  These results underscore the need for further research on FR VQA for HDR and SDR content. The predictions are plotted against the scores obtained for TV1 in Fig.\ref{fig:frpredictions}.

\begin{table*}
\caption{Median SRCC, LCC, and RMSE of FR VQA algorithms on the new LIVE HDRvsSDR Database. Standard deviations are in parentheses. Best results for each TV are in bold.}
\begin{center}
\resizebox{\textwidth}{!}{
\begin{tabular}{|c|c|c|c|c|c|c|c|c|c|} %|c|c|c|c|c|c|c|c|c|c|
\hline
{\textsc{Method}}  &  \multicolumn{3}{|c|}{TV1} & \multicolumn{3}{|c|}{TV2} & \multicolumn{3}{|c|}{TV3} \\  
\hline
 & SRCC$\uparrow$ & PLCC$\uparrow$ & RMSE$\downarrow$ & SRCC$\uparrow$ & LCC$\uparrow$ & RMSE$\downarrow$ & SRCC$\uparrow$ & LCC$\uparrow$ & RMSE$\downarrow$ \\
 \hline
PSNR & 0.1056 & 0.3170 & 18.8984 & 0.0380 & 0.2673 & 19.1349 & 0.1351 & 0.2918 & 17.7954  \\
\hline
 STRRED\cite{strred} & 0.1865 & 0.3228 & 18.8593 & 0.1287 & 0.2691 & 19.1252 & 0.1166 & 0.2331 & 18.0922\\
\hline
SpEED\cite{speed} &  0.3209 & 0.4535 & 17.7103 &  0.2326 & 0.3796 & 17.4674 &  0.1771 & 0.3095 & 16.6531\\
\hline
 MS-SSIM \cite{msssim} & 0.4172 & 0.4391 & 17.9224 & 0.3071 & 0.3502 & 17.3550 & 0.3181 & 0.3544 & 17.3256 \\
\hline
 SSIM \cite{ssim} & 0.4807 & 0.4921 & 17.3459 & 0.4184 & 0.4439 & 17.7940 & 0.3818 & 0.4021 & 17.0344\\
\hline
\textbf{ VMAF \cite{vmaf}} & \textbf{0.5654} & \textbf{0.5850} & \textbf{16.1606} & \textbf{0.5547} & \textbf{0.5810} & \textbf{16.1618} & \textbf{0.5971} & \textbf{0.6057} & \textbf{14.8045} \\
\hline
 STGREED \cite{greed} & 0.4873\scriptsize(0.0908) & 0.5114\scriptsize(0.0643) & 17.3710\scriptsize(1.0645) & 0.4719\scriptsize(0.0902) & 0.4814\scriptsize(0.0635) & 17.3219\scriptsize(1.1705) & 0.4471\scriptsize(0.1191) & 0.4663\scriptsize(0.0758) & 16.1962\scriptsize(1.2141)  \\
\hline
\end{tabular}}
\label{tab:fr_metrics}
\end{center}
\end{table*}

\subsection{HDRPatchMAX}

Motivated by the low performance of existing models, we also designed a new HDR NR-VQA algorithm called HDRPatchMAX. HDRPatchMAX utilizes features that are relevant to SDR and HDR quality, as well as to motion perception. 

\subsubsection{NIQE features}
The first set of features in HDRPatchMAX are the same as those in NIQE~\cite{niqe}. NIQE can be used alone as a blind metric, but we average the 36 features and the distance measure across all frames in each video and use the resulting 37 features for subsequent training. 

\subsubsection{PatchMAX}

The 37 features obtained from NIQE performed strongly on the LIVE HDR, LIVE AQ HDR, and LIVE HDRvsSDR dataset, which prompted us to develop a patch-based feature extraction method. We start by partitioning the frame into non-overlapping patches of size $P\times P$ and computing their Mean Subtracted Contrast Normalization (MSCN) coefficients. The MSCN coefficients $\hat{V}\lbrack i,j,k_0 \rbrack$ of a luma channel of a video frame or patch $V\lbrack i,j,k_0 \rbrack$ are defined as :
\begin{equation}\label{eq:mscn} 
\hat{V}\lbrack i,j ,k_0 \rbrack = \frac{V\lbrack i,j,k_0 \rbrack-\mu\lbrack i,j,k_0 \rbrack}{\sigma\lbrack i,j ,k_0 \rbrack +C},    
\end{equation}
where $i\in 1,2..,M$, $j=1,2..,N$ are spatial indices, $M$ and $N$ are the patch height and width, respectively, the constant $C$ imparts numerical stability, and where
  \begin{equation}\label{eq:mean} 
  \mu\lbrack i,j,k_0 \rbrack = \sum\limits_{m=-L}^{m=L} \sum\limits_{l=-L}^{l=L} w\lbrack m,l\rbrack V\lbrack i+m,j+l,k_0 \rbrack
  \end{equation}
  and
\begin{equation}\label{eq:sigma} 
\begin{split}
\sigma\lbrack i,j,k_0 \rbrack & =( \sum\limits_{m=-L}^{m=L} \sum\limits_{l=-L}^{l=L}    w\lbrack m,l \rbrack (V\lbrack i+m,j+l,k_0 \rbrack\\
& -\mu\lbrack i,j,k_0 \rbrack)^2)^{\frac{1}{2}}
\end{split}
\end{equation}

are the local weighted spatial mean and standard deviation of luma, respectively. The weights $w=\{w[m,l]|m=-L,\dots,L,l=-L,\dots,L\}$ are a 2D circularly-symmetric Gaussian weighting function sampled out to 3 standard deviations and rescaled to unit volume, and $K=L=3$.

The standard deviation that is computed during the MSCN operation is averaged across each patch and is used as a proxy for each patch's contrast. The patches are then divided into three groups based on the  percentile of their standard deviation relative to the standard deviation of the other patches in the frame. High contrast patches are defined as patches having standard deviations above the $(100-T)$\textsuperscript{th} percentile, medium contrast patches are defined as patches with standard deviations between the $T$\textsuperscript{th} and $(100-T)$\textsuperscript{th} percentiles, and low contrast patches are those having standard deviations less than the $T$\textsuperscript{th} percentile. 

The MSCN coefficients of these patches can be reliably modelled as Generalized Gaussian Distributions (GGD), defined as

  \begin{equation}
  g_1(x;\alpha;\beta) = \frac{\alpha}{2\beta \Gamma(\frac{1}{\alpha})} \exp [-(\frac{|x|}{\beta})^\alpha]
  \end{equation}\label{ggd}
where $\Gamma(.)$ is the gamma function:
\begin{equation}
\Gamma(\alpha) = \int_0^\infty t^{\alpha-1} \exp(-t) dt.
\end{equation}
The shape and variance parameters of the best fit to the MSCN coefficients, $\alpha$ and $\beta$, are extracted and used as quality-aware features.

Following this, compute the products of neighboring pairs of pixels in each patch to capture correlations between them as follows:
\begin{align}\label{eq:pair}
\begin{split}
H[i,j,k_0] & = \hat{V}[i,j,k_0]\hat{V}[i,j+1,k_0] \\
V[i,j,k_0] & = \hat{V}[i,j,k_0]\hat{V}[i+1,j,k_0] \\
D_1[i,j,k_0] &= \hat{V}[i,j,k_0]\hat{V}[i+1,j+1,k_0] \\
D_2[i,j,k_0] & = \hat{V}[i,j,k_0]\hat{V}[i+1,j-1,k_0].
\end{split}
\end{align}

These are modelled as following an Asymmetric Generalized Gaussian Distribution (AGGD): 

\begin{equation}\label{eq:aggd}
g_2(x;\nu,\sigma_l^2,\sigma_r^2) = \begin{cases}
\frac{\nu}{(\beta_l+\beta_r)\Gamma (\frac{1}{\nu})} \exp(-(-\frac{x}{\beta_l})^\nu) &  x<0 \\
\frac{\nu}{(\beta_l+\beta_r)\Gamma (\frac{1}{\nu})} \exp(-(\frac{x}{\beta_r})^\nu) &  x>0,
\end{cases} 
\end{equation}
where
\begin{equation}
\beta_l = \sigma_l \sqrt{\frac{\Gamma(\frac{1}{\nu})}{\Gamma(\frac{3}{\nu})}} \quad \mathrm{and} \quad  \beta_r = \sigma_r \sqrt{\frac{\Gamma(\frac{1}{\nu})}{\Gamma(\frac{3}{\nu})}},
\end{equation}
and where $\nu$ controls the shape of the distribution and $\sigma_l$ and $\sigma_r$ control the spread on each side of the mode. The parameters ($\eta,\nu,\sigma_l^2,\sigma_r^2$) are extracted from the best AGGD fit to the histograms of each of the pairwise products in (\ref{eq:pair}), where 
\begin{equation}\label{eq:eta}
\eta = (\beta_r-\beta_l) \frac{\Gamma(\frac{2}{\nu})}{\Gamma(\frac{1}{\nu})}.
\end{equation}

The parameters of the best fit to each patch are extracted as quality features, and are averaged separately for each of the three categories of patches, yielding a total of 54 features. This procedure is performed at two scales to yield a total of 108 features.  In addition to this, we also compute the average temporal standard deviation of these features over every non-overlapping group of five consecutive frames, and use the additional 108 features as spatio-temporal quality features. The patch size $P$ and the percentile threshold $T$ are treated as hyperparameters, and were chosen as $P=20$ and $T=10$ based on the results shown in Table~\ref{tab:results_patchsizeptle}.

There are three important motivations for the use of contrast-separated feature aggregation. Firstly, one of the primary ways HDR differs from SDR is in its representation of local contrast. SDR frames, being limited to 8 bits, cannot represent edges and details as well as HDR frames are. However, feature responses corresponding to this increased contrast visibility may be masked by feature responses from other regions of the frame which may not benefit from the more accurate quantization of HDR. Explicitly separating feature responses by contrast prevents this masking effect. For example, in Fig.~\ref{fig:rchr_hdr_vis}, high contrast regions of an HDR frame are shown in white, medium contrast regions are shown in gray, and low contrast regions are shown in black. In Fig.~\ref{fig:rchr_sdr_vis}, the same is shown for the SDR version of the same frame. In the SDR version, there are underexposed regions in the woman's t-shirt and in the back of the suit worn by the man in the foreground, which are highlighted as low-contrast regions by our proposed contrast-based segmentation. In the HDR version, the same regions are not underexposed and low-contrast regions are instead only located in the plain walls in the background. The statistics of these regions will accordingly be different and quality-aware.

Secondly, NSS modelling suffers in the presence of regions of very low contrast (such as the sky). Prior observations on the Gaussianity of MSCN coefficients of pristine frames are strongly validated by most natural scenes, very smooth areas lacking any texture or detail can present rare exceptions to this natural law. Separating such areas from regions of higher contrast and texture may therefore improve the validity and power of these models.
                                                                           
Thirdly, contrast masking is an important visual phenomenon whereby distortions may become less visible in the presence of high contrast textures, providing additional motivation for separately analyzing regions having different contrasts. The MSCN operation serves to model the contrast-gain masking that occurs in the early stages of the human visual system, while the explicit separation of regions with different contrasts contributes to additional modelling of this effect.

The MSCN coefficients of different patches grouped by their contrasts are plotted in Fig.~\ref{fig:contrastmscn} for a frame from a compressed HDR video, along with the corresponding SDR frame. The coefficients clearly differ and the best GGD fits to these distributions will be accordingly quality-aware.

\subsubsection{HDRMAX}

The HDRMAX feature set~\cite{hdrmax} is extracted by first passing the luma values of each frame through an expansive nonlinearity, first introduced in HDR ChipQA. Values in overlapping windows of size $20 \times 20$ are first linearly rescaled to $[-1,1]$ using the minimum and maximum values in each window. They are then passed through an expansive nonlinearity defined as follows:

\begin{equation}
\label{eq:exp_piecewise}
        f(x;\delta) =   \begin{cases} \exp(\delta x)-1 & x>=0 \\
    1-\exp((-\delta x)) & x<0 \end{cases}
\end{equation}
where $\delta=4$ based on prior experiments~\cite{hdrmax} on the LIVE HDR database.
This nonlinearity amplifies local contrasts by suppressing the middle range of values and by amplifying the extreme ends of the luminance range. Since HDR excels in representing contrasts, this operation enhances endemic compression distortions and forces subsequent feature responses to focus on them. 

\begin{figure*}
     \centering
          \begin{subfigure}[b]{0.3\textwidth}
         \centering
         \includegraphics[width=\textwidth]{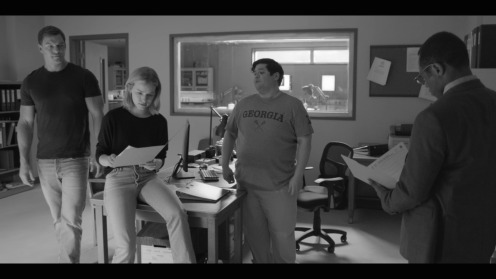}
         \caption{HDR Grayscale}
         \label{fig:rchr_hdr}
     \end{subfigure}
     \hfill
     \begin{subfigure}[b]{0.3\textwidth}
         \centering
         \includegraphics[width=\textwidth]{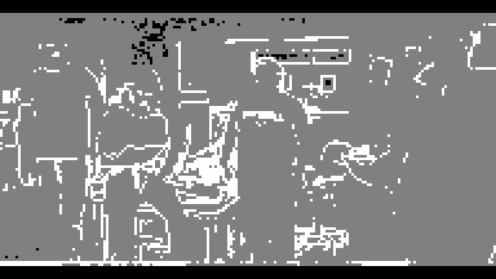}
         \caption{Contrast-based segmentation of HDR}
         \label{fig:rchr_hdr_vis}
     \end{subfigure}
     \hfill
     \begin{subfigure}[b]{0.3\textwidth}
         \centering
         \includegraphics[width=\textwidth]{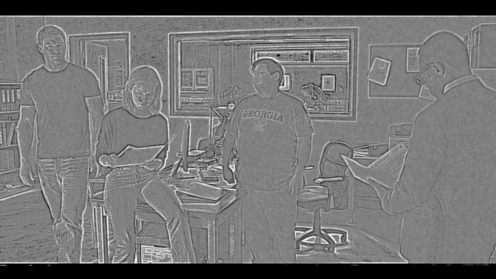}
         \caption{HDR after HDRMAX nonlinearity}
         \label{fig:rchr_hdr_lnl}
     \end{subfigure}
     \\
          \begin{subfigure}[b]{0.3\textwidth}
         \centering
         \includegraphics[width=\textwidth]{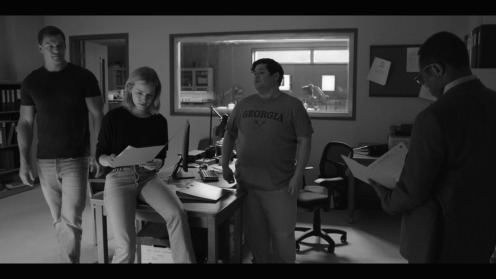}
         \caption{SDR Grayscale}
         \label{fig:rchr_sdr}
     \end{subfigure}
     \hfill
          \begin{subfigure}[b]{0.3\textwidth}
         \centering
         \includegraphics[width=\textwidth]{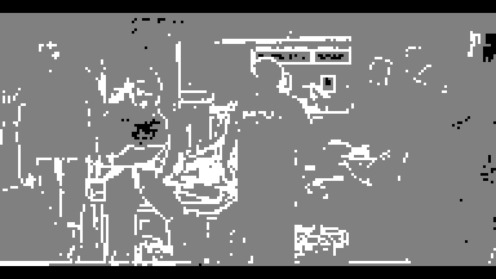}
         \caption{Contrast-based segmentation of SDR}
         \label{fig:rchr_sdr_vis}
     \end{subfigure}
     \hfill
     \begin{subfigure}[b]{0.3\textwidth}
         \centering
         \includegraphics[width=\textwidth]{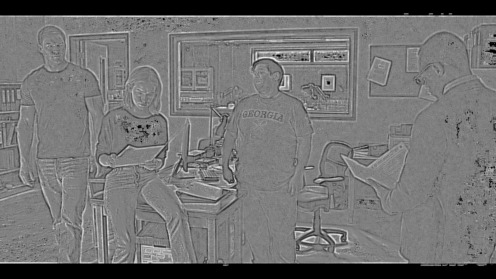}
         \caption{SDR after HDRMAX nonlinearity.}
         \label{fig:rchr_sdr_lnl}
     \end{subfigure}
        \caption{Comparing HDR and SDR.}
        \label{fig:comparison}
\end{figure*}

\begin{figure*}
     \centering
          \begin{subfigure}[b]{0.3\textwidth}
         \centering
         \includegraphics[width=\textwidth]{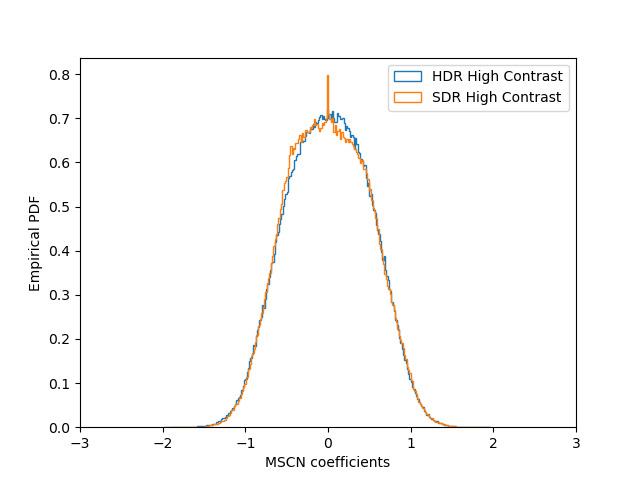}
         \caption{High Contrast Patches}
         \label{fig:mscn_high}
     \end{subfigure}
     \hfill
     \begin{subfigure}[b]{0.3\textwidth}
         \centering
         \includegraphics[width=\textwidth]{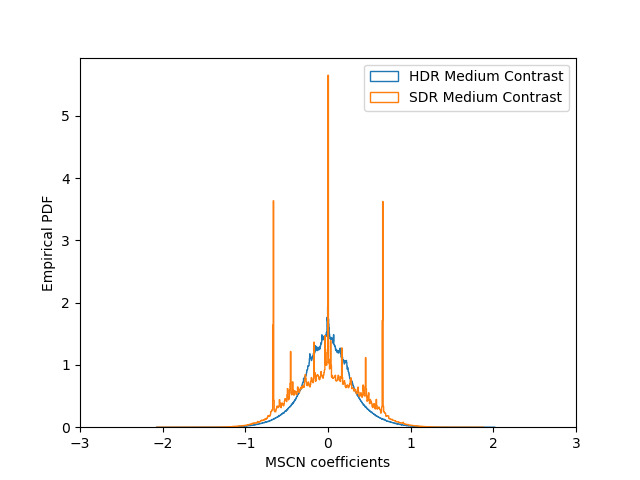}
         \caption{Medium Contrast Patches}
         \label{fig:mscn_med}
     \end{subfigure}
     \hfill
     \begin{subfigure}[b]{0.3\textwidth}
         \centering
         \includegraphics[width=\textwidth]{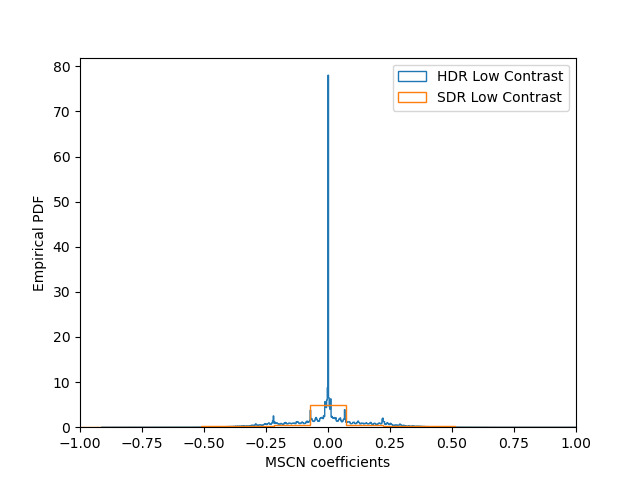}
         \caption{Low Contrast Patches}
         \label{fig:mscn_low}
     \end{subfigure}
     \caption{Comparing HDR and SDR MSCN coefficients of different contrast groups.}
     \label{fig:contrastmscn}
\end{figure*}

In Fig.~\ref{fig:rchr_hdr_lnl} and Fig.~\ref{fig:rchr_sdr_lnl}, for example, the HDRMAX nonlinearity is applied to the frames shown in Fig.~\ref{fig:rchr_hdr} and Fig.~\ref{fig:rchr_sdr}, respectively. Local contrast is amplified in the resulting images throughout the frame. Regions in SDR suffering from underexposure, such as the woman's t-shirt, and overexposure, such as the tubelight in the room behind, present luminance values that are near constant and are hence stretched to extremes with sharp boundaries that highlight their defects. The corresponding regions in the HDR version are smoother, and distortions in those regions highlight HDR's enhanced ability to represent contrast.

The MSCN coefficients of the nonlinearly processed frames are computed and modelled as following a GGD, and their neighboring products are modelled as following an AGGD, yielding 18 features. These features are extracted at two scales. In addition, we also find the average standard deviation of these features on every non-overlapping group of five frames, yielding 72 features.

\subsubsection{Space-Time Chips}

We model spatio-temporal information using Space-Time (ST) Gradient Chips, first introduced in \cite{chipqa0} and further developed in \cite{chipqa}. Gradients carry important information about edges and distortions and ST chips are spatiotemporal slices of the gradient video that are designed to capture temporal information in a quality-aware way. The gradient magnitude of each frame is first computed using a $3\times 3$ Sobel operator. The MSCN coefficients of the gradient magnitude are found using Eqn.~(\ref{eq:mscn}). Following this, a temporal bandpass filter $k(t)$ is applied:
\begin{equation}\label{eq:ht} 
h(t) = t(1-at)\exp (-2at)u(t).
\end{equation}

This simple linear filter models the process of temporal decorrelation that occurrs in the lateral geniculate nucleus (LGN)~\cite{lgn}. The visual signal passes to area V1 from the LGN, where neurons are sensitive to motion over local areas. Motivated by this, spatiotemporal slices of the MSCN coefficients of the gradient video are selected from each spatial location by performing a grid search over six directions, and by selecting the slice having the kurtosis closest to that of a Gaussian.

Specifically, let $\hat{G}[i,j,k]$ denote MSCN coefficients of the gradient magnitudes of video frames with temporal index $k$. For $k=T$, define a block of 5 frames ending at time $T$, denoted $\hat{G}_T = {\hat{G}[i,j,k];k=T-4,\dots T}$. Then, within each space-time volume $\hat{G}_T$, define six 5x5 space-time slices or chips, which intersect all 5 frames, which are separated by an angle of $\frac{\pi}{6}$, and are constrained so that the center of each chip coincides with the center of the volume $\hat{G}_T$, and so that the normal vector of every chip lies on the spatial plane. ST chips that are perpendicular to the direction of motion will capture the motions of objects along that vector, where we tacitly assume that motions of small spatiotemporal volumes is translational. We have previously shown that the natural bandpass statistics of ST chips of high-quality videos that are oriented perpendicular to the local direction of motion reliably follow a Gaussian law, while those pointing in other directions diverge from Gaussianity~\cite{chipqa}. This observation follows from the Gaussianity of MSCN coefficients of pristine frames. Accordingly, in each $5\times 5 \times 5$ volume, we select the $5\times 5$ ST chip that has the least excess kurtosis. The ST Gradient chips coefficients are modelled as following a GGD, and their neighboring pairwise products are modelled as following an AGGD, yielding 18 features. This procedure is repeated at two scales, yielding a total of 36 features.

HDRPatchMAX hence consists of 37 NIQE features, 108 PatchMAX features, 72 HDRMAX features, and 36 ST Gradient Chip features.

\begin{table}
\caption{Effect of varying patch size and percentile thresholds on median SRCC for TV1.}
\begin{center} 
  \begin{tabular}{|l|l|l|}

  \hline
 Patch size $P$ & Percentile threshold $T$ & SRCC  \\ 
\hline
20 & 10 & 0.8347\scriptsize(0.0828)\\
\hline
20 & 20 &  0.8244\scriptsize(0.0902) \\
\hline
20 & 30 & 0.7917\scriptsize(0.1144)\\
\hline
120 & 10 & 0.8035\scriptsize(0.0997)\\
\hline
120 & 20 & 0.8003\scriptsize(0.1061)\\
\hline
120 & 30 & 0.8055\scriptsize(0.1035)\\
\hline
240 & 10 & 0.7838\scriptsize(0.1197)\\
\hline
240 & 20 & 0.7853\scriptsize(0.1123)\\
\hline
240 & 30 & 0.7838\scriptsize(0.1197)\\
\hline
\end{tabular}
\label{tab:results_patchsizeptle}
\end{center}
\end{table}

%Motion features are defined by modelling the statistics of frame differences. Adjacent frames are subtracted and the MSCN coefficients of the frame differences are computed. The frame-differences are divided into patches, and the average value of each patch of the frame-difference is computed as a proxy for motion. Motion can mask distortions and regions with high motion will have different video statistics from regions with low or no motion. It is therefore beneficial to analyze the statistics of regions differently based on their motion. Similar to how contrast is used to segment the frame, the patches with an average frame-difference value above the top 75th percentile are classified as high-motion, patches with a an average frame-difference value between the 25th and 75th percentile are classified as medium motion patches, and patches with an average frame-difference value less than the 25th percentile are classified as low motion patches. BRISQUE features are computed for each patch of frame-differences separately and averaged across each group. This procedure is performed at two scales to yield an additional 36 features.

\subsection{No-Reference Video Quality Assessment}

We evaluated RAPIQUE~\cite{rapique}, BRISQUE~\cite{brisque}, TLVQ~\cite{tlvqm}, VBLIINDS~\cite{vbliinds}, ChipQA~\cite{chipqa}, and HDR ChipQA~\cite{hdrchipqa}, as well as HDRPatchMAX, on the new LIVE HDRvsSDR Database. Again, we report the SRCC, PLCC, and RMSE metrics. The algorithms were trained using a Random Forest Regressor. We also tested the Support Vector Regressor but chose the Random Forest Regressor based on its better performance. The videos are separated into training and test sets with an 80:20 ratio such that videos of the same content appeared in the same set. Cross-validation was performed over the training set to select the best hyperparameters (the number of estimators and the number of features) for the random forest. The random forest with the best hyperparameters was then fitted to the training set and evaluated on the test set. This procedure was repeated 100 times with different randomized train-test splits, with the median results reported along with the standard deviations in Table~\ref{tab:results_nr}. HDRPatchMAX outperformed the other NR VQA algorithms on all the televisions. The parameters $P$ and $T$ were chosen based on the performance of the PatchMAX set of features as $P$ and $T$ were varied, as reported in Table~\ref{tab:results_patchsizeptle}.
\begin{table*}
\caption{Median SRCC, LCC, and RMSE of NR VQA algorithms on LIVE HDRvsSDR. Standard deviations are in parentheses. Best results for each TV are in bold.}
\begin{center} 
\resizebox{\textwidth}{!}{
  \begin{tabular}{|c|c|c|c|c|c|c|c|c|c|}
  \hline
   Dataset &
      \multicolumn{3}{c|}{TV1} & \multicolumn{3}{c|}{TV2} & \multicolumn{3}{c|}{TV3} \\
  \hline
 & SRCC$\uparrow$ & PLCC$\uparrow$ & RMSE$\downarrow$ & SRCC$\uparrow$ & PLCC$\uparrow$ & RMSE$\downarrow$ & SRCC$\uparrow$ & PLCC$\uparrow$ & RMSE$\downarrow$\\ 
\hline
RAPIQUE & 0.4814\scriptsize(0.1315) & 0.5769\scriptsize(0.1099) & 17.1953\scriptsize(2.1580) & 0.5163\scriptsize(0.1382) & 0.5713\scriptsize(0.1332) & 17.5951\scriptsize(1.7378) & 0.5116\scriptsize(0.1455) & 0.5279\scriptsize(0.1402) & 16.4634\scriptsize(2.0299)\\
\hline
BRISQUE & 0.6923\scriptsize(0.1339) & 0.7337\scriptsize(0.1221) & 15.6301\scriptsize(2.7714) & 0.7355\scriptsize(0.1009) & 0.7736\scriptsize(0.0820) & 13.2101\scriptsize(2.4794)& 0.7285\scriptsize(0.1094) & 0.7583\scriptsize(0.0920) & 12.9504\scriptsize(2.5819)  \\
\hline
TLVQM & 0.7340\scriptsize(0.1022) & 0.7865\scriptsize(0.0873) & 12.9524\scriptsize(2.3336) & 0.7422\scriptsize(0.0876) & 0.7976\scriptsize(0.0698) & 12.4863\scriptsize(2.0532)& 0.7734\scriptsize(0.0929) & 0.8087\scriptsize(0.0747) & 11.7250\scriptsize(1.9711)\\
\hline
ChipQA & 0.7396\scriptsize(0.1040) & 0.7810\scriptsize(0.0918) & 13.0415\scriptsize(2.3269) &  0.7450\scriptsize(0.1052) & 0.7933\scriptsize(0.0822) & 11.9339\scriptsize(1.8891) & 0.7943\scriptsize(0.1038) & 0.7912\scriptsize(0.0915) & 12.2313\scriptsize(2.0476)\\
\hline  
HIGRADE & 0.7610\scriptsize(0.1196) & 0.7928\scriptsize(0.1076) & 12.6439\scriptsize(2.7403) & 0.7950\scriptsize(0.1189) & 0.8246\scriptsize(0.0970) & 11.4478\scriptsize(2.3973) & 0.7497\scriptsize(0.1074) & 0.7679\scriptsize(0.0898) & 12.5681\scriptsize(2.2225) \\
\hline 
VBLIINDS & 0.7895\scriptsize(0.0867) & 0.8060\scriptsize(0.0793) & 13.4819\scriptsize(2.3147) & 0.8005\scriptsize(0.0825) & 0.8216\scriptsize(0.0698) & 12.0408\scriptsize(2.1253)& 0.8052\scriptsize(0.0833) & 0.8085\scriptsize(0.0665) & 11.2490\scriptsize(2.0545)\\
\hline
VIDEVAL & 0.7913\scriptsize(0.0815) & 0.8210\scriptsize(0.0722) & 11.8524\scriptsize(2.0402) & 0.8014\scriptsize(0.0862) & 0.8248\scriptsize(0.0702) & 11.8614\scriptsize(2.0022) & 0.8252\scriptsize(0.0719) & 0.8275\scriptsize(0.0648) & 11.0899\scriptsize(1.8907)\\
\hline
HDRChipQA & 0.7941\scriptsize(0.0954) & 0.8129\scriptsize(0.0913) & 11.9236\scriptsize(2.4481) &  0.8116\scriptsize(0.0774) & 0.8209\scriptsize(0.0767) & 12.0703\scriptsize(2.2669) &  0.8112\scriptsize(0.0899) & 0.8135\scriptsize(0.0867) & 11.5474\scriptsize(2.3609)\\
\hline
HDRPatchMAX & \textbf{0.8586\scriptsize(0.0723)} & \textbf{0.8524\scriptsize(0.0654)} & \textbf{10.8200\scriptsize(2.0941)} & \textbf{0.8436\scriptsize(0.0714)} & \textbf{0.8588\scriptsize(0.0636)} & \textbf{11.1900\scriptsize(2.0942)} & \textbf{0.8494\scriptsize(0.0691)} & \textbf{0.8551\scriptsize(0.0627)} & \textbf{10.2514\scriptsize(2.0489)} \\
\hline
\end{tabular}}
\label{tab:results_nr}
\end{center}
\end{table*}

\begin{figure*}
     \centering
     \begin{subfigure}[b]{0.24\textwidth}
         \centering
         \includegraphics[width=\textwidth]{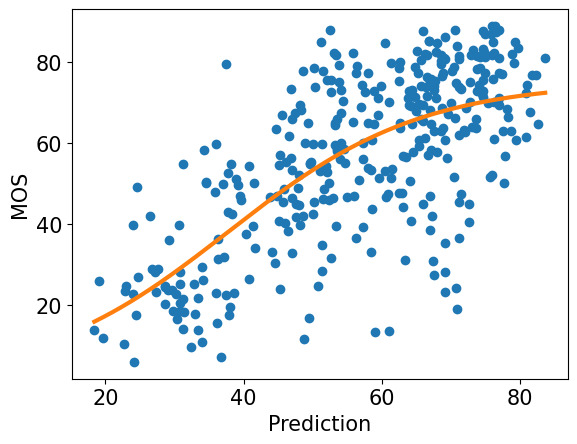}
         \caption{BRISQUE}
         \label{fig:brisque}
     \end{subfigure}
               \hfill
     \begin{subfigure}[b]{0.24\textwidth}
         \centering
         \includegraphics[width=\textwidth]{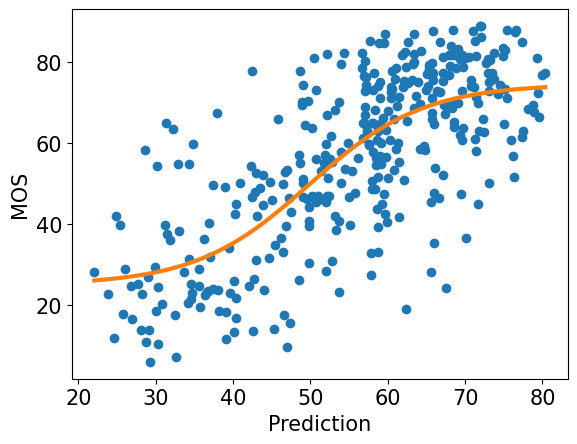}
         \caption{ChipQA}
         \label{fig:chipqa}
     \end{subfigure}
          \hfill
     \begin{subfigure}[b]{0.24\textwidth}
         \centering
         \includegraphics[width=\textwidth]{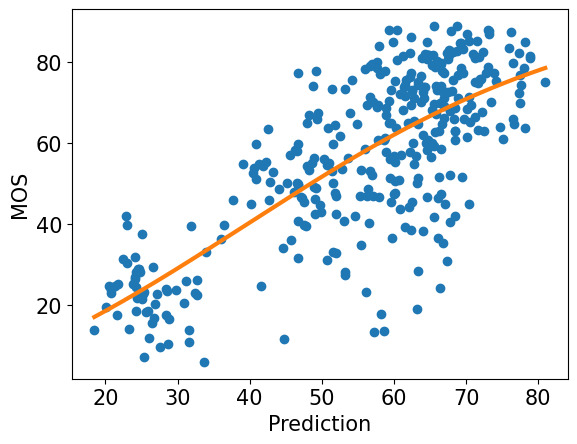}
         \caption{TLVQM}
         \label{fig:tlvqm}
     \end{subfigure}
           \hfill
          \begin{subfigure}[b]{0.24\textwidth}
         \centering
         \includegraphics[width=\textwidth]{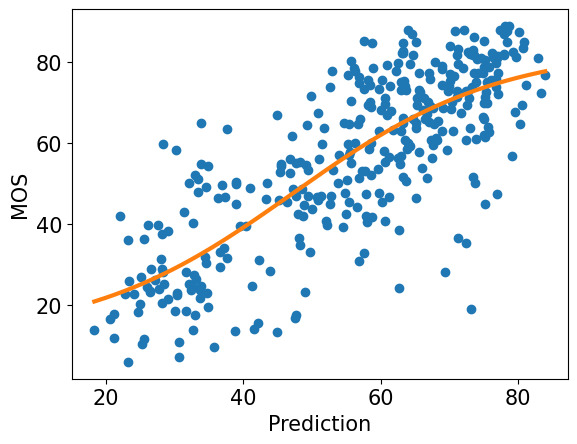}
         \caption{VBLIINDS}
         \label{fig:vbliinds}
     \end{subfigure}
      \\
               \begin{subfigure}[b]{0.24\textwidth}
         \centering
         \includegraphics[width=\textwidth]{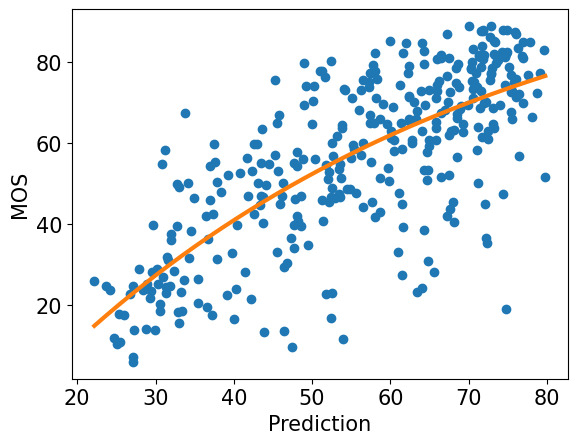}
         \caption{HIGRADE}
         \label{fig:higrade}
     \end{subfigure}
     \hfill
     \begin{subfigure}[b]{0.24\textwidth}
         \centering
         \includegraphics[width=\textwidth]{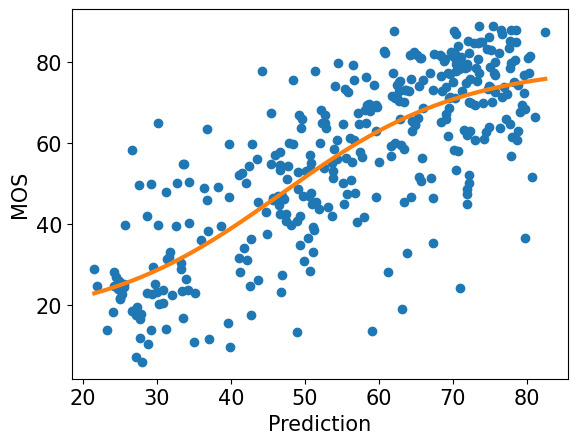}
         \caption{VIDEVAL}
         \label{fig:videval}
     \end{subfigure}
          \hfill
     \begin{subfigure}[b]{0.24\textwidth}
         \centering
         \includegraphics[width=\textwidth]{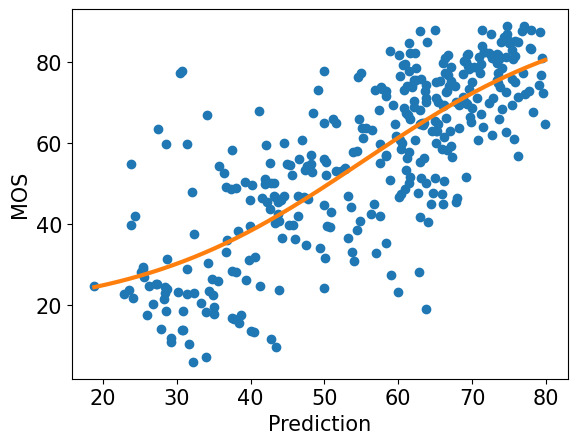}
         \caption{HDR ChipQA}
         \label{fig:hdrchipqa}
     \end{subfigure}
    \hfill
          \begin{subfigure}[b]{0.24\textwidth}
         \centering
         \includegraphics[width=\textwidth]{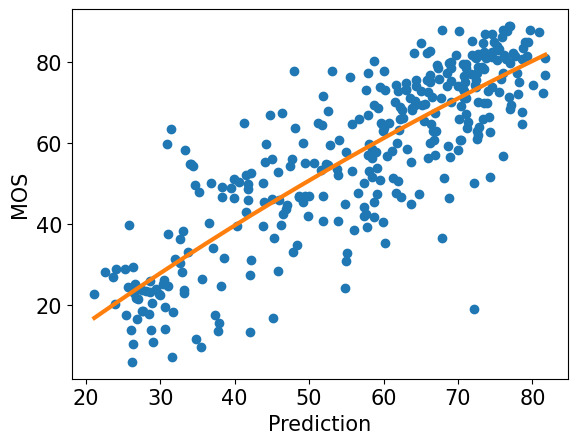}
         \caption{HDRPatchMAX}
         \label{fig:hdrpatchmax}
     \end{subfigure}
\caption{Scatter plots of MOS against NR VQA predictions with logistic fit in orange.}
        \label{fig:nrpredictions}
\end{figure*}

The scatter plots of the predictions made by the NR VQA algorithms against the MOS obtained on the videos shown on TV1 are shown in Fig.~\ref{fig:nrpredictions}. The scatter plots were created by plotting the MOS against the mean quality predictions produced by the NR VQA algorithms on each video in the test set over the 100 train-test splits, using the logistic fit in Eqn.~\ref{eq:logistic} shown in orange. HDRPatchMAX's predictions are better aligned with MOS and have a more linear fit than the other algorithms.

\begin{figure*}
    \centering
    \includegraphics[width=\textwidth]{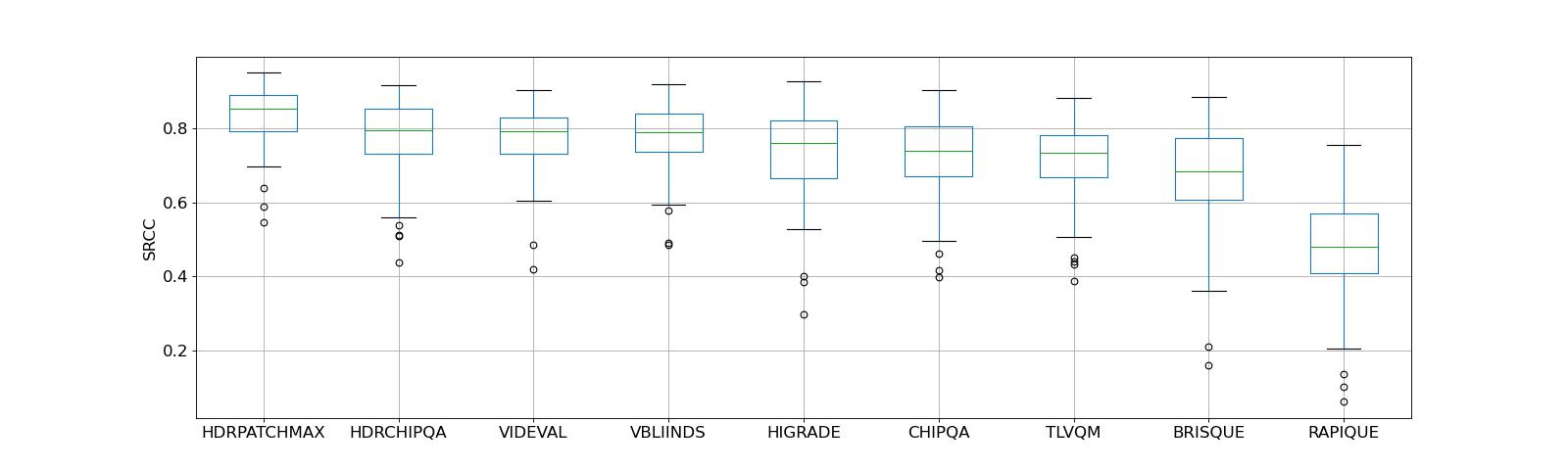}
    \caption{Boxplot of SRCCs obtained by NR VQA algorithms for 100 train-test splits}
    \label{fig:boxplot}
\end{figure*}

We also show a boxplot of the SRCCs obtained over the 100 train-test splits by the NR VQA algorithms on TV1 in Fig.~\ref{fig:boxplot}. HDRPatchMAX has a higher median SRCC and a tighter spread of values than the other algorithms.

We also evaluated each feature set in HDRPatchMAX individually, and report the results in Table~\ref{tab:individual} for TV1. The PatchMAX set of features obtained the highest performance and individually exceeded the performance of other NR VQA algorithms. We also conducted an ablation study with HDRPatchMAX, removing each feature set and evaluating the performance of the rest of the feature sets on TV1, and report the results in Tab.~\ref{tab:ablation}. Despite the strong individual performance of PatchMAX, its removal had minimal impact on the algorithm's performance, indicating the robustness and complementary nature of the other feature sets. Similar results were observed on TV2 and TV3.

\begin{table}
\caption{Individual feature performance on TV1.}
\begin{center} 
  \begin{tabular}{|l|l|l|l|l|}

  \hline
 Feature  & SRCC$\uparrow$ & PLCC$\uparrow$ & RMSE$\downarrow$  \\ 
\hline
ST Grad. Chips & 0.5872\scriptsize(0.1380) & 0.6264\scriptsize(0.1347) & 17.4190\scriptsize(2.5293) \\
\hline 
HDRMAX & 0.7688\scriptsize(0.0647) & 0.7900\scriptsize(0.0682) & 13.9499\scriptsize(1.8542)\\
\hline
NIQE & 0.7829\scriptsize(0.0896) & 0.8057\scriptsize(0.0809) & 13.5136\scriptsize(2.3201)\\
\hline
PatchMAX &  0.8347\scriptsize(0.0828) & 0.8376\scriptsize(0.0794) & 12.7068\scriptsize(2.0619) \\
\hline
\end{tabular}
\label{tab:individual}
\end{center}
\end{table}

\begin{table}
\caption{Ablation study on TV1.}
\begin{center} 
  \begin{tabular}{|l|l|l|l|l|}
  \hline
 Feature set \textbf{removed} & SRCC$\uparrow$ & PLCC$\uparrow$ & RMSE$\downarrow$  \\ 
\hline
PatchMAX & 0.8553\scriptsize(0.0719) & 0.8567\scriptsize(0.0661) & 11.9267\scriptsize(2.0255)\\
\hline
ST Grad. Chips  & 0.8524\scriptsize(0.0721) & 0.8579\scriptsize(0.0665) & 12.0564\scriptsize(1.9838)\\
\hline 
NIQE & 0.8428\scriptsize(0.0720) & 0.8405\scriptsize(0.0715) & 12.3666\scriptsize(1.9113)\\
\hline
HDRMAX  & 0.8334\scriptsize(0.0777) & 0.8375\scriptsize(0.0709) & 12.5534\scriptsize(1.9670) \\
\hline
\end{tabular}
\label{tab:ablation}
\end{center}
\end{table}

The random forest models trained on the data from TV1, TV2, and TV3, can be considered as separate models for OLED displays, QLED displays, and LED displays respectively. However, we also conducted a separate experiment where each TV was assigned a numerical value (1 for TV1, 2 for TV2, and 3 for TV3) that was concatenated onto each NR VQA algorithm as an additional feature, and trained on the scores obtained from all the display devices, treating videos shown on different display devices as different stimuli with different feature vectors. The results are shown in Table~\ref{tab:alltvs}. HDRPatchMAX once again is the leading performer.

\begin{table}
\caption{Results on all display devices using TV indices as additional features.}
\begin{center} 
  \begin{tabular}{|l|l|l|l|l|}
  \hline
Algorithm & SRCC$\uparrow$ & PLCC$\uparrow$ & RMSE$\downarrow$  \\ 
\hline
RAPIQUE & 0.5106\scriptsize(0.1306) & 0.5789\scriptsize(0.1106) & 16.0880\scriptsize(1.9977)\\
\hline
BRISQUE & 0.7091\scriptsize(0.1221) & 0.7500\scriptsize(0.0926) & 13.1656\scriptsize(2.2399)\\
\hline 
ChipQA & 0.7319\scriptsize(0.1046) & 0.7748\scriptsize(0.0884) & 12.4513\scriptsize(2.0950) \\
\hline 
HIGRADE & 0.7673\scriptsize(0.1105) & 0.7937\scriptsize(0.0947) & 12.3120\scriptsize(2.3021)\\
\hline
TLVQM & 0.7491\scriptsize(0.0988) & 0.7957\scriptsize(0.0817) & 12.0173\scriptsize(2.1908) \\
\hline
HIGRADE & 0.7673\scriptsize(0.1105) & 0.7937\scriptsize(0.0947) & 12.3120\scriptsize(2.3021) \\ 
\hline
VBLIINDS & 0.7850\scriptsize(0.0812) & 0.8098\scriptsize(0.0600) & 11.4007\scriptsize(1.7672) \\
\hline
HDRChipQA & 0.7864\scriptsize(0.0894) & 0.8050\scriptsize(0.0878) & 11.3890\scriptsize(2.3069)\\
\hline 
VIDEVAL & 0.7969\scriptsize(0.0812) & 0.8243\scriptsize(0.0733) & 11.2577\scriptsize(1.9316)\\
\hline 
\textbf{HDRPatchMAX} & \textbf{0.8367\scriptsize(0.0732)} & \textbf{0.8437\scriptsize(0.0605)} & \textbf{10.4734\scriptsize(1.9913)} \\
\hline 
\end{tabular}
\label{tab:alltvs}
\end{center}
\end{table}

\subsection{Results on combined databases}
{We evaluated the NR and FR VQA algorithms on the combined LIVE HDR, LIVE AQ HDR, and LIVE HDRvsSDR databases consisting of 891 HDR videos and 175 SDR videos. The Samsung Q90T (TV2) was used for the LIVE HDR and LIVE AQ HDR studies, so the scores from the LIVE HDR and LIVE AQ HDR databases were mapped to the scores obtained from from TV2 in the LIVE HDRvsSDR database using the procedure described in Section~\ref{section:combiningdata}, and then used as the ground truth for evaluation. The results are presented in Table~\ref{tab:combined_fr_metrics} for FR metrics and Table~\ref{tab:combined_nr_metrics} for NR metrics. HDRPatchMAX was the best performer on the combined dataset among NR VQA models while VMAF was the best performer among FR VQA models.}

\begin{table}
\caption{Median SRCC, LCC, and RMSE of FR VQA algorithms on the combined databases. Standard deviations are in parentheses. Best results are in bold.}
\begin{center}
\begin{tabular}{|c|c|c|c|c|c|c|c|c|c|} %|c|c|c|c|c|c|c|c|c|c|
\hline
{\textsc{Method}}  & SRCC$\uparrow$ & PLCC$\uparrow$ & RMSE$\downarrow$\\  
\hline
PSNR & 0.3208 & 0.4777 & 16.5218 \\
\hline
 SSIM \cite{ssim} & 0.3618 & 0.3726 & 17.4523 \\
\hline
 MS-SSIM \cite{msssim} & 0.3693 & 0.3794 & 16.5699  \\
\hline
 SpEED\cite{speed} &  0.3235 & 0.4237 & 17.0150 \\
\hline
 STRRED\cite{strred} & 0.3588 & 0.3921 & 17.3007\\
\hline
\textbf{VMAF\cite{vmaf}} &  \textbf{0.6710} & \textbf{0.6930} & \textbf{13.5578} \\
\hline
 STGREED \cite{greed} & 0.6182\scriptsize(0.0687) & 0.6150\scriptsize(0.0641) & 14.8379\scriptsize(1.0856) \\
\hline
\end{tabular}
\label{tab:combined_fr_metrics}
\end{center}
\end{table}

\begin{table}
\caption{Median SRCC, LCC, and RMSE of FR VQA algorithms on the combined databases. Standard deviations are in parentheses. Best results are in bold.}
\begin{center} 
  \begin{tabular}{|l|l|l|l|l|}
  \hline
Algorithm & SRCC$\uparrow$ & PLCC$\uparrow$ & RMSE$\downarrow$  \\ 
\hline
RAPIQUE & 0.5684\scriptsize(0.0828) & 0.5440\scriptsize(0.0736) & 16.0103\scriptsize(0.9990)\\
\hline
BRISQUE & 0.6551\scriptsize(0.0621) & 0.6374\scriptsize(0.0560) & 14.5969\scriptsize(1.0516)\\
\hline 
ChipQA & 0.7319\scriptsize(0.1046) & 0.7748\scriptsize(0.0884) & 12.4513\scriptsize(2.0950) \\
\hline 
HIGRADE & 0.7673\scriptsize(0.1105) & 0.7937\scriptsize(0.0947) & 12.3120\scriptsize(2.3021)\\
\hline
TLVQM &  0.6997\scriptsize(0.0601) & 0.6916\scriptsize(0.0547) & 13.7720\scriptsize(1.0247)\\
\hline
HIGRADE &  0.6780\scriptsize(0.0736) & 0.6670\scriptsize(0.0665) & 13.9237\scriptsize(1.2181)\\ 
\hline
VBLIINDS & 0.7501\scriptsize(0.0572) & 0.7394\scriptsize(0.0524) & 12.8156\scriptsize(1.2017) \\
\hline
VIDEVAL & 0.7588\scriptsize(0.0559) & 0.7526\scriptsize(0.0529) & 12.4759\scriptsize(1.0645)\\
\hline 
HDRChipQA & 0.7699\scriptsize(0.0537) & 0.7702\scriptsize(0.0484) & 12.2328\scriptsize(1.1006)\\
\hline 
\textbf{HDRPatchMAX} & \textbf{0.8064\scriptsize(0.0516) }& \textbf{0.8065\scriptsize(0.0445)} & \textbf{11.2318\scriptsize(0.9964)} \\
\hline 
\end{tabular}
\label{tab:combined_nr_metrics}
\end{center}
\end{table}

\section{Conclusion}

We presented the first ever study on comparing HDR and SDR videos of the same content encoded at different bitrates and resolutions on different display devices. Our study shows that despite HDR's theoretical capabilities over SDR, its perceptual quality depends heavily on the display device used in practice. We also evaluated several NR and FR VQA algorithms on the new database, and presented a novel NR VQA algorithm called HDRPatchMAX that exceeds the current state-of-the-art on this database. We hope that this spurs research on the modelling of display devices in VQA algorithms as well as for optimal bitrate ladders for streaming.

\section*{Acknowledgment}

This research was sponsored by a grant from Amazon.com, Inc., and by grant number 2019844 for the National Science Foundation AI Institute for Foundations of Machine Learning (IFML). The authors also thank the Texas Advanced Computing Center (TACC) at The University of Texas at Austin for providing HPC resources that have contributed to the research results reported in this paper. URL: http://www.tacc.utexas.edu. 

\ifCLASSOPTIONcaptionsoff
  \newpage
\fi



\bibliographystyle{IEEEtran}
\bibliography{hdr_tip}




%\newpage


%\vfill


\end{document}